\documentclass[aps,prx,reprint,superscriptaddress,showpacs]{revtex4-2}
\usepackage{graphicx}
\usepackage{subfigure}
\usepackage{dcolumn}
\usepackage{bm}
\usepackage{hyperref}
\usepackage{indentfirst}
\usepackage{braket}
\usepackage{amsmath}
\usepackage{amssymb}
\usepackage[normalem]{ulem}
\hypersetup{colorlinks=true, citecolor=blue, urlcolor=blue, linkcolor=blue}

\begin{document}

\title{Topological Chiral and Nematic Superconductivity by Doping Mott Insulators  on Triangular Lattice}

\author{Yixuan Huang}
\affiliation{Department of Physics and Astronomy, California State University, Northridge, California 91330, USA}

\author{D. N. Sheng}
\email{donna.sheng1$@$csun.edu}
\affiliation{Department of Physics and Astronomy, California State University, Northridge, California 91330, USA}

\date{\today}

\begin{abstract}
The mechanism of the unconventional topological superconductivity (TSC) remains a long-standing issue. We investigate the quantum phase diagram  of the extended $t$-$J$-$J_{\chi}$ model including spin chiral interactions on triangular lattice based on the state-of-the-art density matrix renormalization group simulations. We identify  distinct classes of superconducting phases characterized by nonzero topological Chern numbers $C=1$ and $2$, and a nematic d-wave superconducting phase with a zero Chern number. The TSC states are shown to emerge from doping either a magnetic insulator or chiral spin liquid,  which opens new opportunities  for experimental discovery. In addition, we further classify the $C=2$ class of TSC phases into  an isotropic and a nematic TSC phases,  and present evidence of continuous quantum phase transitions from the nematic TSC phase to both isotropic TSC and nematic d-wave phases. These results provide new insight into the  mechanism of TSC with an emphasis on the role played by hole dynamics, which changes spin background and drives a topological phase transition at a hole doping level around $3\%$ upon doping a magnetic insulator to enable the emergence of the TSC.
\end{abstract}


\maketitle

\section{Introduction}
\label{introduction}

There have been intensive studies of the canonical models for  strongly correlated systems, the two-dimensional (2D) Hubbard and $t$-$J$ models, and their generalized versions since the discovery of high-$T_c$  cuprate superconductivity (SC)~\cite{anderson1987resonating,lee2006doping,keimer2015quantum,proust2019remarkable,ogata2008t,fradkin2015colloquium,arovas2022hubbard}.
At the strong coupling limit, these models host different  Mott insulating states varying from  magnetic insulators to spin liquids~\cite{balents2010spin}. 
Understanding the interplay of conventional orders, spin liquid physics and unconventional SC in doped  Mott insulators  is one of the central
challenges of condensed matter physics.
A large  body of work on unconventional SC is connected to  the  original proposal of the resonating valence bond theory~\cite{anderson1987resonating} that doping  Mott insulators might naturally lead to SC~\cite{senthil2005cup,lee2006doping,ogata2008t,fradkin2015colloquium,weng1999mean,song2021doping}.
Lacking well-controlled analytical solutions in 2D with strong couplings, unbiased numerical studies play an important role in establishing the quantum phases in such models.  
Along this direction, exciting progress has been made in understanding the emergence of SC  and its interplay with spin fluctuations and charge stripes by doping the antiferromagnetic Mott state on the square lattice based on extensive numerical simulations~\cite{white2009pairing,corboz2014competing,PhysRevX.5.041041,jiang2019superconductivity,qin2020,zheng2017stripe,jiang2018superconductivity,dodaro2017intertwined,jiang2020ground,PhysRevLett.127.097002,gong2021robust,jiang2021ground}, which is relevant to cuprate SC.
In particular, more recent density matrix renormalization group (DMRG)~\cite{white1992density} studies have established robust SC for extended $t$-$J$ and Hubbard models on square lattice with next-nearest-neighbor hoppings suggesting the importance of tuning hole dynamics to enhance SC~\cite{jiang2019superconductivity,PhysRevLett.127.097002,gong2021robust,jiang2021ground}.

Mott insulating states on triangular lattice offer another exciting playground and challenges for their distinct interplay between geometric frustrations, lattice rotational symmetry,  and quantum fluctuations~\cite{baskaran2003electronic,kumar2003superconductivity,wang2004doped,motrunich2004,raghu2010superconductivity,chen2013unconventional,venderley2019density,jiang2020topological,song2021doping,jiang2021superconductivity,zhu2022doped,arovas2022hubbard,gannot2020SU,szasz2020chiral,chen2021quantum,peng2021gapless,wietek2021,aghaei2020efficient}. 
On the experimental side, the SC state observed in Na$_x$CoO$_2$·yH$_2$O might be a $d+id$-wave topological superconductivity (TSC) state which breaks time-reversal symmetry~\cite{takada2003superconductivity,zhou2008nodal}. More recently, different twisted transition metal dichalcogenide (TMD) moir$\acute{e}$ systems have been discovered to be quantum simulators of the Hubbard model~\cite{wu2018hubbard,Tang2020},  which are promising systems  hosting  correlated insulators and topological superconductors~\cite{an2020interaction,schrade2021nematic,scherer2021n}.

Theoretically, the Kalmeyer-Laughlin (KL) chiral spin liquid (CSL)~\cite{kalmeyer1987equivalence,wen1989chiral} has been identified among the phase boundaries of different competing magnetic ordered states~\cite{bauer2014,he2014,gong2014,gong2017global,wietek2017chiral} for frustrated spin systems or near the Mott transition for the Hubbard model~\cite{szasz2020chiral,chen2021quantum} on triangular lattice. Whether doping a KL CSL can generally lead to the TSC~\cite{kalmeyer1987equivalence,wen1989chiral,laughlin1988,lee1989,jiang2017holon,jiang2020topological,zhu2022doped,song2021doping}
remains an open question. A recent theoretical study~\cite{song2021doping} suggests that doping a KL CSL may naturally lead to a chiral metal while a topological $d+id$-wave SC represents a more nontrivial scenario requiring the internal gauge flux to be adjusted with the hole doping level.
Indeed, unbiased numerical simulations have found a possible chiral metal by doping the CSL identified at half filling of the triangular Hubbard model~\cite{szasz2020chiral, zhu2022doped}.
The  nontrivial  example of identifying the topological $d+id$-wave SC by doping the KL CSL comes from  the study of the  $t$-$J$-$J_{\chi}$ model~\cite{jiang2020topological} with strong three-spin chiral interactions.  The  topological class of the observed TSC state~\cite{jiang2020topological}  characterized by a finite  integer quantized spin Chern number~\cite{read2000paired, senthil1999spin} and chiral Majorana edge modes has not been revealed. Crucially, the driving mechanism  for the emergence of the TSC  remains to be identified, which may require extensive exploration in the parameter space by tuning relevant hopping parameters and interactions~\cite{gong2017global,wietek2017chiral,jiang2020topological}. Given the fact that CSLs often emerge near the boundaries between different magnetically ordered states~\cite{gong2017global,wietek2017chiral}, related open questions  naturally arise including what the interplay is between the TSC and conventional orders or fluctuations, and whether distinct unconventional SC states can emerge  by doping  different magnetically ordered states.   

To address these open issues, we study the quantum phase diagram and focus on the emergent unconventional SC in the  extended $t$-$J$-$J_{\chi}$ model on triangular lattice  based on large-scale DMRG simulations~\cite{white1992density}.  
By tuning the ratios of the next-nearest- and  nearest-neighbor hoppings ($t_{2}/t_{1}$) and Heisenberg spin couplings ($J_{2}/J_{1}$) in the presence of  three spin chiral interactions ($J_{\chi}$)~\cite{jiang2020topological}, 
we identify different superconducting phases
including  distinct $d+id$-wave TSC phases that are characterized by nonzero topological Chern numbers  $C=1$ and $2$, and a nematic SC phase with a d-wave pairing symmetry breaking lattice rotational symmetry and $C=0$. Furthermore, 
for weaker $J_{\chi }$ the C=2 phases include isotropic and nematic TSC phases  with a continuous quantum phase transition between them.
We demonstrate that these  topological and  nematic d-wave SC states have robust power-law decaying  pairing correlations in the form of Luther-Emery liquid~\cite{luther1974backward}  on wider cylinders, which may lead to different superconducting states in 2D. 
The TSC can be induced by either doping a magnetically ordered state or CSL, which provides a new opportunity for experimental discovery of unconventional TSC. We also demonstrate  the important role played by hole dynamics, which  can drive a topological phase transition upon doping a $120^{\circ}$ antiferromagnetic (AFM) state at a hole doping level $\delta\approx 3\%$,  enabling the  TSC to emerge.  Furthermore, the nematic SC with $C=0$  
can emerge from either doping the CSL or magnetic ordered states~\cite{gong2017global}, suggesting  the rich interplay between unconventional SC and spin background.

The rest of the paper is organized as follows. In Sec.~\ref{model_method}, we introduce the extended $t$-$J$-$J_{\chi}$ model on a triangle lattice, the DMRG method, and the topological characterization for the SC states through spin flux insertion. Its quantum phase diagram is presented in Sec.~\ref{phase_diagram}, containing different TSC phases and a nematic d-wave SC phase. We demonstrate their distinct topological Chern numbers (Sec.~\ref{flux_insertion}), the quasi-long-range order in SC pairing correlations (Sec.~\ref{SC_correlation}), and the pairing symmetries (Sec.~\ref{pairing_symmetry}) to  characterize these  phases. In Sec.~\ref{symmetry_evolutions} we focus on the quantum phase transitions by tuning the ratios of $t_{2}/t_{1}$ and $J_{2}/J_{1}$, with Sec.~\ref{pairing_order_evolution} showing the evolution of SC pairing correlations, Sec.~\ref{phase_transition} addressing the nature of quantum phase transitions among different phases, and Sec.~\ref{spin_correlations} showing the evolution of spin correlations. The summary and discussions are presented in Sec.~\ref{summary}.

\begin{figure}
\centering
\includegraphics[width=1\linewidth]{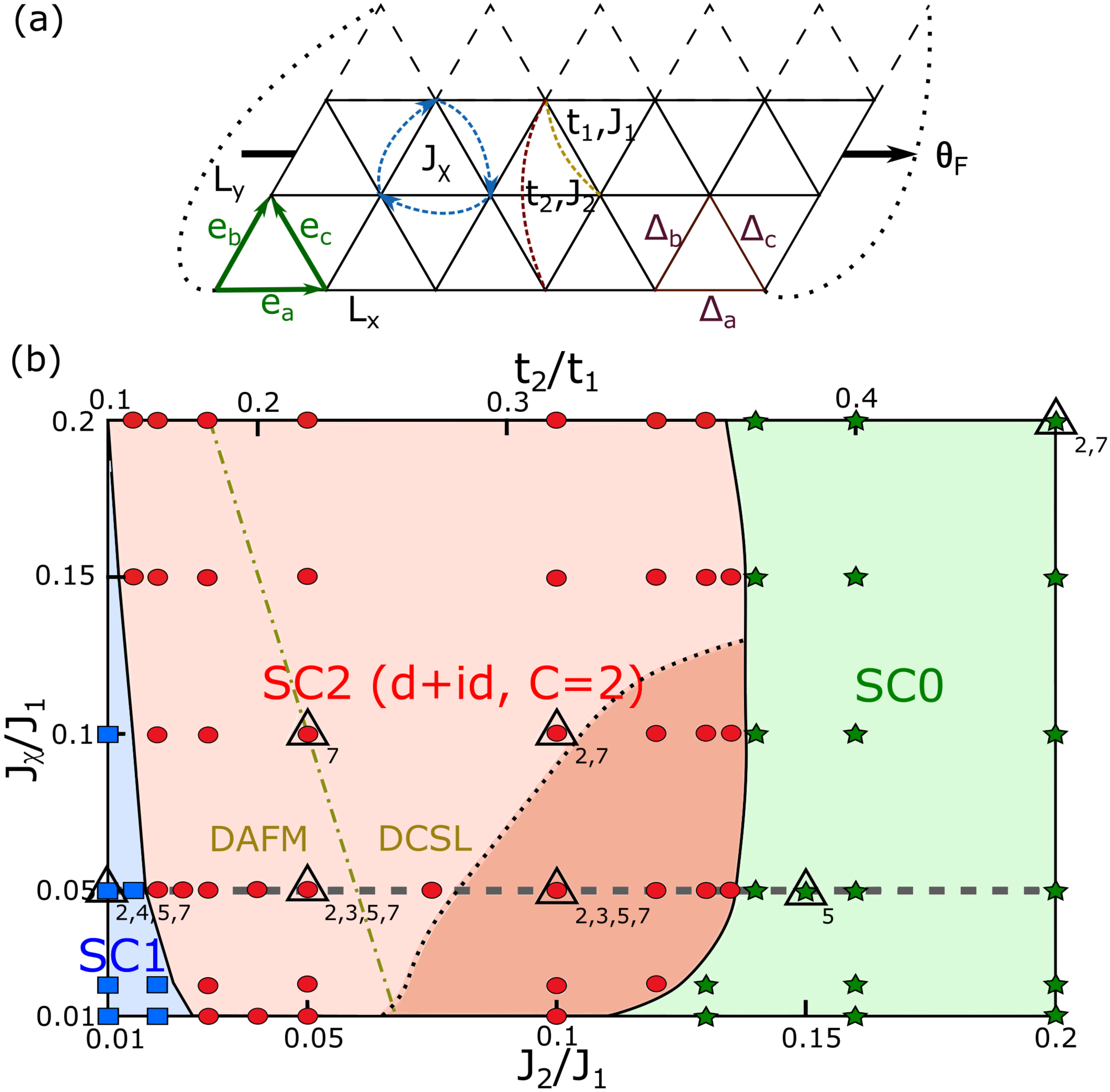}
\caption{Quantum phase diagram. (a) Schematic illustration of the extended $t$-$J$-$J_{\chi}$ model on triangle lattice with the nearest-neighbor and the next-nearest-neighbor hoppings $t_{1}$ and $t_{2}$ and Heisenberg exchange $J_{1}$ and $J_{2}$ interactions, as well as the three-spin chiral interactions $J_{\chi}$. The dashed lines indicate  the periodic boundary condition. (b) The quantum phase diagram obtained on  $L_{y}=6$ cylinders  based on the Chern number simulations. For  $0.01 \leq J_{2}/J_{1}, J_{\chi}/J_{1} \leq 0.2$   and doping level  $\delta=1/12$, we identify distinct classes of SC phases  labeled as SC1, SC2, and SC0 from left to right characterized  by their spin Chern number $C$. The SC2 regime represents two phases, the isotropic TSC (lighter red region) and nematic TSC  (darker red region) phases. Different symbols represent parameter points studied
with the DMRG methods.  The triangles mark  points  presented in the paper with the lower indices representing the indices of figures.  A scan of the Chern number, SC pairing symmetry and energy/entropy along the horizontal dashed line is used in Figs.~\ref{Fig_flux_insertion}(c) ,~\ref{Fig_phase_transition}(d), and~\ref{Fig_nature_transiton}, respectively. The main feature of the phase diagram is essentially the same for other doping levels $\delta=1/24-1/8$.}
\label{Fig_phase_diagram}
\end{figure}

\begin{figure}
\centering
\includegraphics[width=1\linewidth]{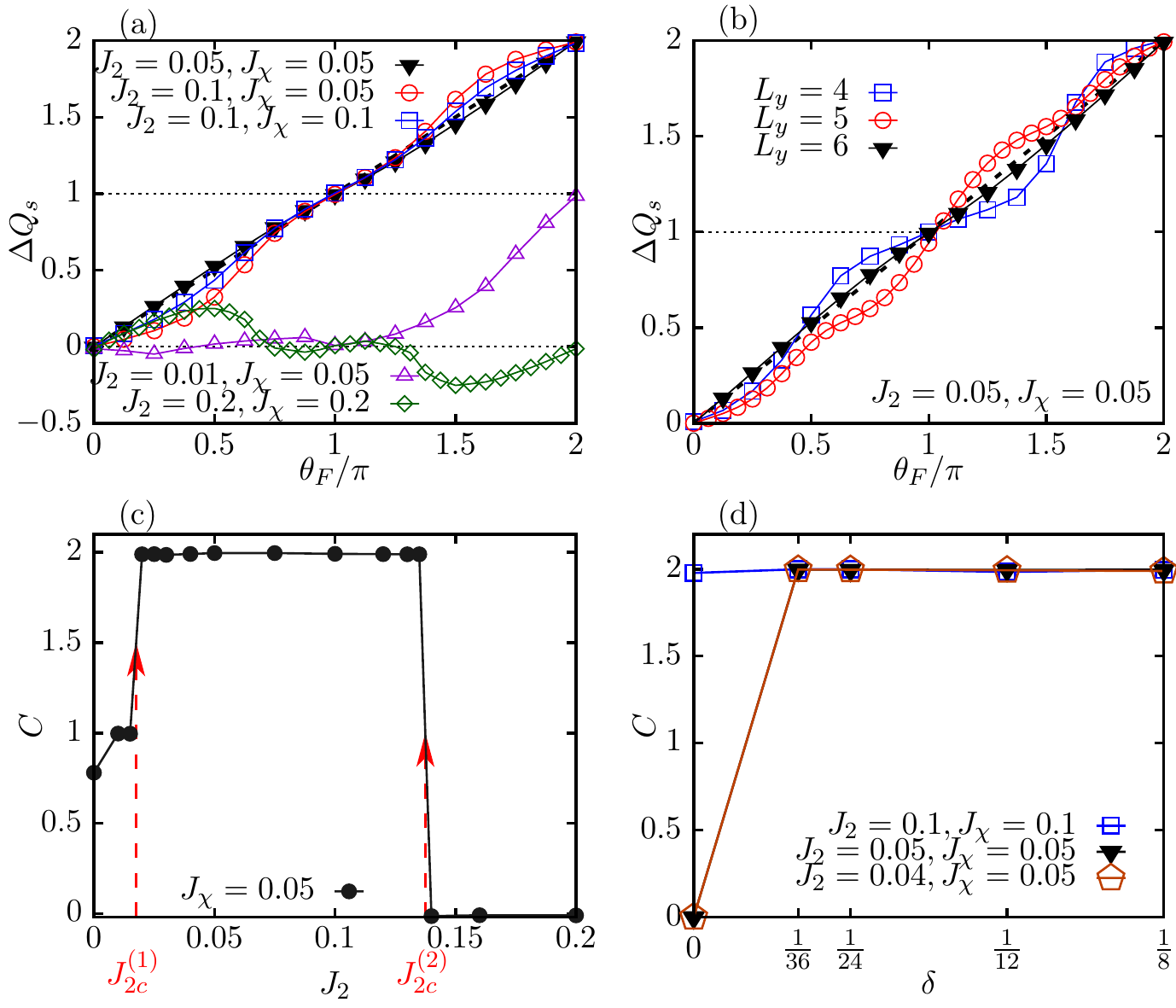}
\caption{The pumped spin by inserting flux and spin Chern number. (a) The pumped spin $\Delta Q_{s}$ with adiabatically inserted flux $\theta_{F}$ is shown for different SC phases  on the $L_{y}=6$ cylinder. For the three parameter points in the SC2 phase the measured spin pumping after inserting one flux quantum $\Delta Q_s|_0^{2\pi}=1.993, 1.983, 1.990$, respectively, indicating that the error bar is around 1\% from the exactly quantized value $\Delta Q_s|_0^{2\pi}=2$ for different systems. (b) $\Delta Q_s$ versus $\theta_F$ for $J_{2}=J_{\chi}=0.05$ inside SC2 phase on $L_{y}=4, 5, 6$ cylinders. (c) The evolution of Chern number $C$ with varying $J_{2}$ at $J_{\chi }=0.05$ on $L_{y}=6$. $C=\Delta Q_s|_0^{2\pi}$ is obtained after inserting one flux quantum $\theta_F=0\rightarrow 2\pi$.  SC1-SC2 and SC2-SC0 phase transitions take place at $J_{2}=J_{2c}^{(1)},J_{2c}^{(2)}$, respectively, where  the  Chern number jumps.  The doping level is $\delta=1/12$ for (a)-(c). (d) $C$ versus $\delta$ for parameter points with different undoped parent states (e.g. DAFM or DCSL) on $L_{y}=6$.}
\label{Fig_flux_insertion}
\end{figure}

\section{Model and method}
\label{model_method}
We study the extended $t$-$J$-$J_{\chi}$ model that is defined as 
\begin{eqnarray}
\label{eq1}
H &=& -\sum\limits_{\left \{ ij \right \},\sigma
}t_{ij}(\widehat{c}^{\dagger }_{i,\sigma }\widehat{c}_{j,\sigma }+H.c.) + \sum\limits_{ \left \{ ij \right \} } J_{ij}(\widehat{\boldsymbol{S}}_{i}\cdot \widehat{\boldsymbol{S}}_{j}-\frac{1}{4}\widehat{n}_{i}\widehat{n}_{j})  \nonumber \\ &+& J_{\chi
}\sum\limits_{\{ ijk \} \in \bigtriangledown / \bigtriangleup  }\widehat{\boldsymbol{S}}_{i}\cdot (
\widehat{\boldsymbol{S}}_{j}\times \widehat{\boldsymbol{S}}_{k}),
\end{eqnarray} 
where $\widehat{c}_{i,\sigma}^{\dagger}$ is the electron creation operator on site $i$ with spin index $\sigma=\pm 1$, $\widehat{\boldsymbol{S}}_{i}$ is the spin-$\frac{1}{2}$ operator and  $\widehat{n}_{i}=\sum_{\sigma}\widehat{c}_{i,\sigma}^+\widehat{c}_{i,\sigma}$ is the  electron number operator.
We consider the  nearest-neighbor and the next-nearest-neighbor hoppings $t_{1}$ and $t_{2}$ as well as Heisenberg couplings $J_{1}$ and $J_{2}$, supplemented by the three-spin chiral interactions $J_{\chi}$ on every elementary triangle as illustrated in Fig.~\ref{Fig_phase_diagram}(a). The chiral interaction can be generated from the Hubbard model with an external magnetic field~\cite{jiang2020topological}.
We set $J_{1}=1$ as the unit of energy, $t_{1}=3$, $J_{2}/J_{1}=(t_{2}/t_{1})^{2}$, and focus on the regime of $0<J_{2}, J_{\chi} \leq 0.2$ with hole doping level $\delta \leq 1/8$, which is the optimal doping region for unconventional SC~\cite{jiang2020topological,gong2021robust}.

To obtain the ground state of the Hamiltonian in Eq.~(\ref{eq1}), we apply the  DMRG method with U(1)$\times $SU(2) for charge and spin symmetries~\cite{McCulloch2007} on cylinder systems with an open boundary condition along the axis ($e_{a}$ or $x-$) direction and a periodic boundary condition along the circumferential ($e_{b}$ or $y-$)  direction, as illustrated in Fig.~\ref{Fig_phase_diagram}(a). The number of sites is $N=L_{x}\times L_{y}$ where  $L_{x}$ and $L_{y}$ denote the lengths in these two directions, respectively. 
The number of electrons $N_{e}$ is related to the doping level $N_{e}/N=1-\delta$.
We keep up to $M=12000$ SU(2) spin multiplets [equivalent to about $m=36000$ U(1) states] with  truncation error  $\epsilon \sim  10^{-6}$, which leads to  accurate results (see Supplemental Sec. II~\cite{SuppMaterial} for details). We develop a topological characterization for the SC  states through the spin flux insertion by adiabatically evolving the ground state as a function of a twisted boundary phase $\theta_F$ based on the method established for CSL and fractional quantum Hall systems~\cite{gong2014,grushin2015characterization}. The flux adds a spin-dependent phase factor to the electron hoppings $\widehat{c}_{i,\sigma}^{\dagger }\widehat{c}_{j,\sigma}\rightarrow e^{i\sigma\theta_{F}}\widehat{c}_{i,\sigma}^{\dagger}\widehat{c}_{j,\sigma}$ if $j\rightarrow i$ crossing the $y$ boundary from the top (see Fig.~\ref{Fig_phase_diagram}(a)), and similarly couples to  the spin flip terms~\cite{gong2014}. In this type of calculation, SU(2) symmetry is broken by the spin flux and we use U(1)$\times $U(1) symmetries with bond dimensions up to $m=8000$ for accurate results due to the robustness of the topologically protected spin pumping (see Supplemental Sec. I~\cite{SuppMaterial} for more details).

\section{Quantum phase diagram}
\label{phase_diagram}
At half filling (with no doping $\delta=0$), the Hamiltonian in Eq.~(\ref{eq1}) reduces to the Heisenberg $J_{1}$-$J_{2}$-$J{\chi}$ model~\cite{gong2017global}. 
In the small $J_{2}$ regime ($J_{1}=1$), the $120^{\circ}$ AFM order survives up to $J_{2}\approx 0.07$ at $J_{\chi}=0$, which
smoothly extends to the nonzero $J_{\chi}$ regime. The intermediate $J_{2}$ regime is dominated by the CSL, which separates from the AFM order by the dash-dotted line as shown in Fig.~\ref{Fig_phase_diagram}(b) obtained from Refs.~\cite{ wietek2017chiral,gong2017global}. Through extensive DMRG simulations of topological Chern numbers  and SC pairing correlations on $L_{y}=$4-6 cylinders for hole-doped systems, we establish a quantum phase diagram in the parameter space $0<J_{2}, J_{\chi} \leq 0.2$ for doping $\delta=1/12$, with three distinct classes of superconducting phases stabilized by  small $J_{2}, J_{\chi} \geq  0.01$~\cite{note1} as shown in Fig.~\ref{Fig_phase_diagram}(b). These SC phases are characterized by  different topological spin Chern numbers. At small $J_{2}$ we find a topological chiral $d+id$-wave SC phase with spin Chern number $C = 1$ (labeled as SC1) by doping the AFM  (DAFM) state. 
In the intermediate $J_{2}$ regime, another class of topological $d+id$-wave SC phases emerges  characterized by a quantized $C = 2$ (SC2), which can be induced by doping either the AFM state or the CSL (DCSL) as illustrated in Fig.~\ref{Fig_phase_diagram}(b). 
The SC2 class is further divided into an isotropic TSC phase and a nematic TSC phase breaking rotational symmetry.
Interestingly,  the nematic TSC state is an analog state
of the recently revealed nematic fractional quantum Hall effect~\cite{haldane2011geometrical,you2014theory, yang2017anisotropic,regnault2017evidence}. 
At larger $J_{2}$, the SC phase has a d-wave symmetry with anisotropic pairing correlations  breaking the lattice rotational symmetry and $C=0$ (SC0) indicating a topologically-trivial SC phase.  The SC0 phase  belongs to the same quantum phase as the nematic d-wave SC identified for an extended $t$-$J$ model~\cite{jiang2021superconductivity} with time-reversal symmetry. The phase  diagram is essentially the same for other doping levels $\delta=$1/24-1/8 with small shifts in phase boundaries (e.g., at $\delta=1/8$,  $\Delta J_{2c}^{(1)}\approx  0.01\sim 0.02$ where $J_{2c}^{(1)}$ denotes the critical $J_{2}$ between SC1 and SC2). We also find that the previously revealed $d+id$-wave SC state (at $J_{\chi}=0.4$ and $J_{2}=t_{2}=0.0$)~\cite{jiang2020topological} has $C=2$ sitting near the phase boundary of the SC2  phase.
\subsection{Topological Chern number characterization  through flux insertion}
\label{flux_insertion}
The nonzero  Chern number characterizes the topological nature of the $d+id$-wave superconductors~\cite{read2000paired,senthil1999spin}, which also identifies the number of chiral Majorana edge modes. 
We determine  the spin Chern number through the spin pumping with inserting flux $\theta _{F}$ into the cylinder, as illustrated in Fig.~\ref{Fig_phase_diagram}(a).  The net spin with nonzero $S_z$ accumulates near the boundaries of the cylinder as the flux adiabatically increases, while the total $S_{z}=0$ for the ground state at different $\theta_F$. We use a small step for the increase of the flux $\theta_F \rightarrow \theta_F+\Delta \theta_F$ with $\Delta \theta_F=2\pi/16$.
A finite Chern number~\cite{gong2014} can be obtained from the total spin  pumping $C=\Delta Q_{s}|_0^{2\pi}= (n_{\uparrow}-n_{\downarrow})|_0^{2\pi}$  measured at the left boundary  at $\theta _{F}=0$ and $2\pi$, where $n_{\sigma}$ is the accumulated charge near the boundary with spin $\sigma$.  We directly measure the pumped spin  for each $\theta_F$ from the reduced density matrix by calculating the sum $Q_{s}=\sum_{\alpha}\lambda _{\alpha}(n_{\uparrow,\alpha}-n_{\downarrow,\alpha})$, where  $\lambda _ {\alpha }$ is the eigenvalue  and $\alpha$ the eigenstate of the reduced density matrix~\cite{grushin2015characterization}, and  $n_{\uparrow,\alpha}$ ($n_{\downarrow,\alpha}$) is the particle number of the  $\alpha$ state with up (down) spin. Because the inserted flux breaks SU(2) symmetry, we use infinite DMRG with U(1)$\times $U(1) symmetries with a large unit cell that is commensurate with the doping level (see Supplemental Sec. I~\cite{SuppMaterial}). 

\begin{figure}
\centering
\includegraphics[width=1\linewidth]{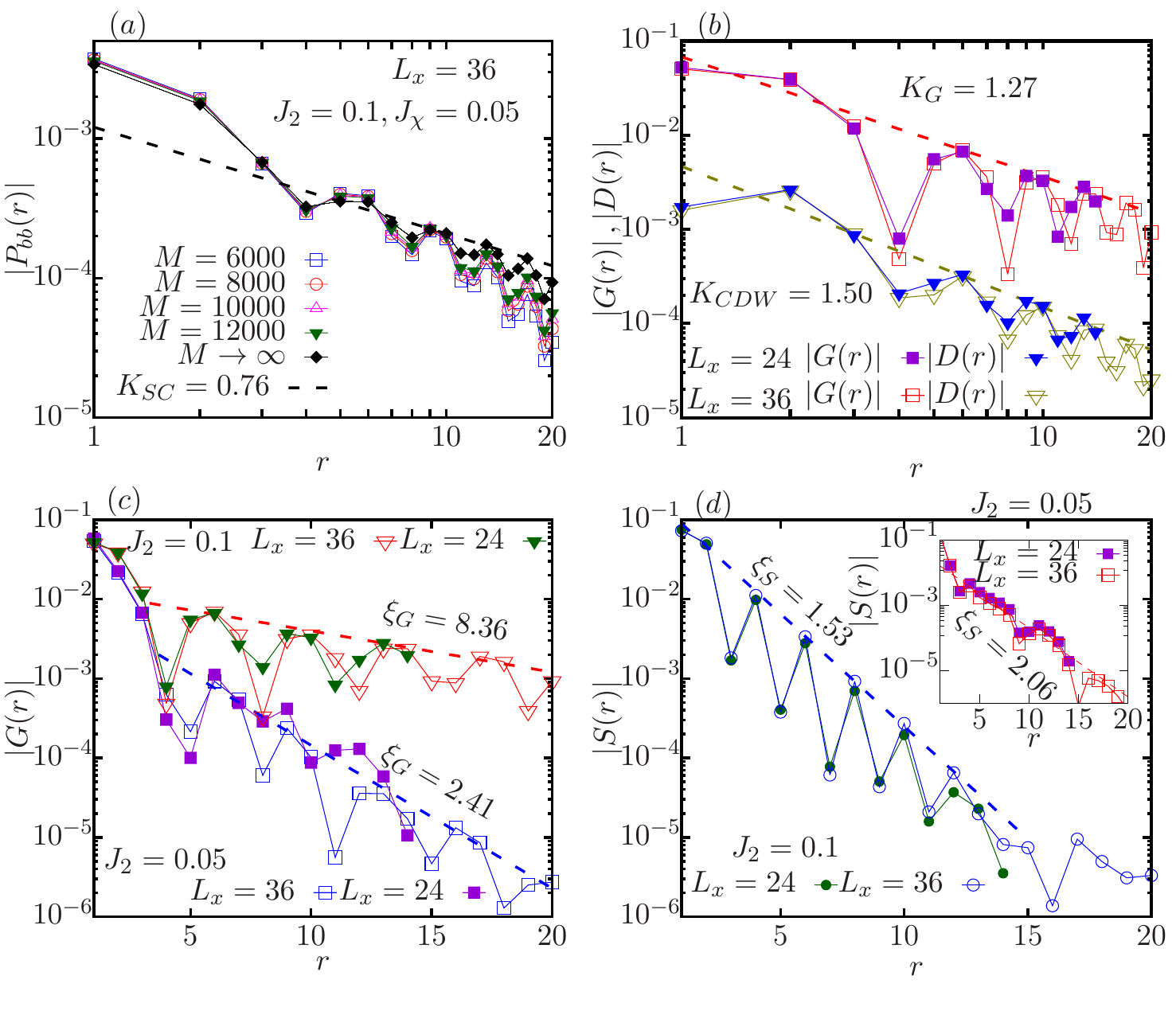}
\caption{Pairing and other correlations in SC2. (a) The SC pairing correlations for different  bond dimensions $M$ at $J_{2}=0.1,J_{\chi}=0.05$ for $N=36\times 6$ system. $\mathbf{r}$ is chosen along the $x$ direction $\mathbf{r}=(r,0)$ and the reference point is $\mathbf{r}_{0}=(L_{x}/4,y_{0})$ to avoid the boundary effect (the results are independent of $y_{0}$ because of the translational invariant along the $y$ direction). The  straight-line fit in the log-log plot of the extrapolated data in the infinite $M$ limit follows a power-law behavior. (b) The density-density correlation $\left |D(r)\right |$ and single-particle correlation $\left |G(r)\right |$ which are fit by the power-law relation for $N=24\times 6$ and $36\times 6$, at $J_{2}=0.1,J_{\chi}=0.05$.
(c) The comparison of $\left |G(r)\right |$ at $J_{2}=0.05$ and $0.1$ with  the same $J_{\chi}=0.05$, which demonstrates the fast growing of the correlation length $\xi_G$ with the increase of $J_{2}$. (d) The exponential decay of the spin correlations $\left |S(r)\right |$ in the main figure and its inset for the same parameters as in (c), where $r > 15$ data points are  ignored in the fitting  because their values are comparable to the numerical truncation error. The doping level is $\delta=1/12$. The obtained fitting exponents or correlation lengths have error bars around 0.02, except for the one in the inset of (d) which is around 0.06.}
\label{Fig_SC_correlation}
\end{figure}

We show examples of the flux insertion and the resulting Chern numbers for systems with $\delta=1/12$ in Fig.~\ref{Fig_flux_insertion}(a). For three parameter points inside the SC2 phase on the $L_{y}=6$ system, the $\Delta Q_{s}$ increases almost linearly with $\theta _{F}$ indicating uniform Berry curvature~\cite{sheng2006}, and there is $\Delta Q_{s}\approx 2.0$ net spin  pumped to  the boundary after the threading of one flux quantum ($\theta_F=0\rightarrow 2\pi$). This corresponds to the quantized Chern number $C=2$, which remains the same on various $L_{y}=4$, $5$, and $6$ as shown in Fig.~\ref{Fig_flux_insertion}(b). The pumping rate becomes more uniform with the increase of $L_{y}$, indicating the increased robustness of the topological quantization for larger systems.   In contrast, at $J_{2}=0.01,J_{\chi}=0.05$ inside the SC1 phase, we find no linear relation between $\Delta Q_{s}$ and $\theta _{F}$ which indicates the nonuniform Berry curvature versus boundary phase $\theta_F$~\cite{sheng2006}. The measured $\Delta Q_{s}\approx 0.983$, indicating around one  net spin  pumped  with the insertion of one flux quantum and $C = 1$. The Chern number becomes nonquantized extended to the $J_{2}=0$ limit  as shown in Fig.~\ref{Fig_flux_insertion}(c) signaling gapless low-energy excitations. We believe the large  variance of the Berry curvature versus $\theta_F$ for the SC1 phase may indicate a  topological state with gapless excitations at small $J_{2}$ consistent with an early proposal for a
topological superconducting state for the Na$_x$CoO$_2$·yH$_2$O system based on variational simulations~\cite{zhou2008nodal}. In the SC0 phase at larger $J_{2}=J_{\chi }=0.2$, we find $\Delta Q_{s}\approx -0.01$ which confirms $C = 0$ for a topologically trivial SC state. Thus, the phase transitions between the three phases can be characterized by jumps of topological Chern number $C$ with varying $J_{2}$ at a fixed $J_{\chi}=0.05$ as illustrated in Fig.~\ref{Fig_flux_insertion}(c). 

Since the undoped $(\delta=0)$ parent state of the SC2 phase contains both the AFM state and  CSL, a natural question is how the Chern number evolves with the doping level. As demonstrated in Fig.~\ref{Fig_flux_insertion}(d), for two points in the DAFM regime at $J_{2}=0.04,0.05$ and $J_{\chi}=0.05$, $C$ jumps from 0 to 2 at  a small doping of $\delta=1/36$ and remains quantized at $C=2$ for larger $\delta$, which demonstrates a doping-induced topological quantum phase transition. On the contrary, in the DCSL regime at $J_{2}=J_{\chi}=0.1$, $C=2$ for $\delta=$0-1/8 which shows a robust Chern number quantization from the parent CSL to the topological $d+id$-wave SC. This is consistent with the fact that the KL CSL is a bosonic $\nu =1/2$  fractional quantum Hall  state~\cite{kalmeyer1987equivalence,wen1989chiral,read2000paired}, which is equivalent to the $C=2$ topological order for fermionic systems where the phase space is enlarged by a factor of 4 in the definition of the Chern number~\cite{hu2015variational} with a doubled flux period  for the Hamiltonian to be invariant. The exact quantization $C=2$ of the SC2 phase indicates that doped holes can indeed adjust the internal flux with the hole doping level to realize a bosonic integer quantum Hall effect for holons~\cite{song2021doping}. 

\subsection{Quasi-long-range order in superconducting pairing correlations}
\label{SC_correlation}
To explore the superconducting nature of the system,  we focus on  the dominant spin singlet pairing correlations $P_{\alpha \beta }(\mathbf{r})=<\widehat{\Delta} ^{\dagger }_{\alpha }(\mathbf{r}_{0})\widehat{\Delta} _{\beta }(\mathbf{r}_{0}+\mathbf{r})>$, where the pairing operator $\widehat{\Delta} _{\alpha }(\mathbf{r})=(\widehat{c}_{\mathbf{r}\uparrow}\widehat{c}_{\mathbf{r}+e_{\alpha }\downarrow}-\widehat{c}_{\mathbf{r}\downarrow}\widehat{c}_{\mathbf{r}+e_{\alpha }\uparrow})/\sqrt{2}$  with $\alpha = a,b,c$, representing different nearest-neighboring bonds as illustrated in Fig.~\ref{Fig_phase_diagram}(a). 

\begin{figure}
\centering
\includegraphics[width=1\linewidth]{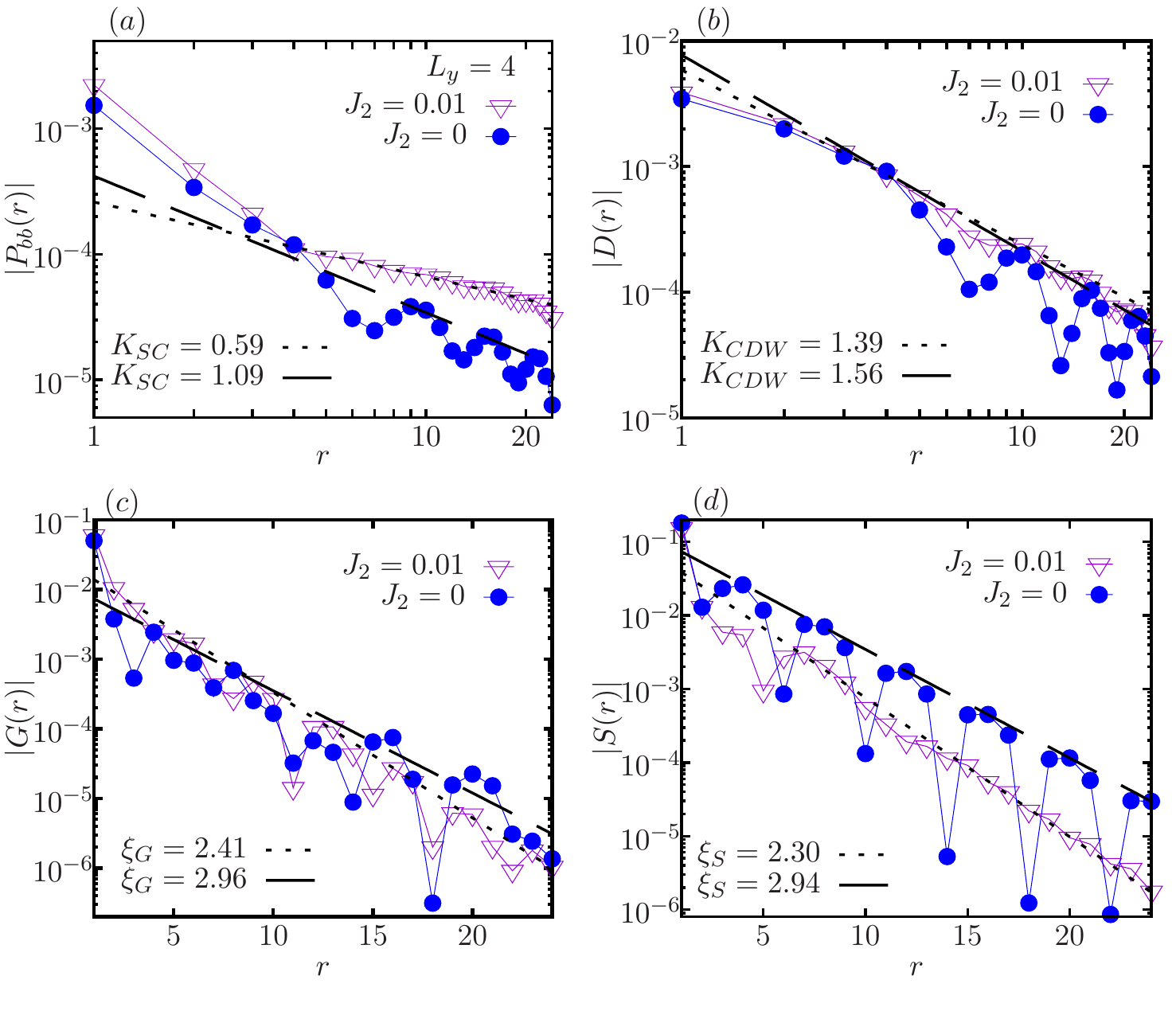}
\caption{Pairing and other correlations at $J_{\chi}=0.05$ for smaller $J_{2}$ on $L_{y}=4$. The results are converged with a large bond dimension $M=10000$. (a) The SC pairing correlations. $\mathbf{r}$ is chosen along x-direction $\mathbf{r}=(r,0)$ and the reference point is $\mathbf{r}_{0}=(L_{x}/4,L_{y}/2)$ to avoid the boundary effect. The straight-line fit in the log-log plot follows a power-law behavior, with  exponents of $1.09$ and $0.59$ for $J_{2}=0$ and $0.01$, respectively. (b) The density-density correlations $\left |D(r)\right |$ which are fit by a power-law relation, with  exponents of $1.56$ and $1.39$ for $J_{2}=0$ and $0.01$, respectively. (c) The single particle correlations $\left |G(r)\right |$ which are fit by an exponential decay with  correlation lengths of $2.96$ and $2.41$ for $J_{2}=0$ and $0.01$, respectively. (d) The spin correlations $\left |S(r)\right |$ which are fit by an exponential decay with  correlation lengths of $2.94$ and $2.30$ for $J_{2}=0$ and $0.01$, respectively. The doping level is $\delta=1/12$. The obtained fitting exponents or correlation lengths have  error bars around 0.03.}
\label{Fig_SC1}
\end{figure}

We first give an example of  the SC pairing correlations in the SC2 regime as shown in Fig.~\ref{Fig_SC_correlation}(a), where the magnitude of pairing correlations at longer distance for two $b$ bonds (along the $y$ direction) $\left | P_{bb}(r) \right |$  increases gradually as the DMRG bond dimension increases from $M=6000$ to $12000$ at $J_{2}=0.1,J_{\chi}=0.05$ on $N=36\times 6$ system.
Because the DMRG represents the ground state in the matrix product form~\cite{schollwock2011density} with finite bond dimensions, the scaling to $M\rightarrow \infty$ is needed to identify the true nature of long-distance correlations for wider cylinders. 
Using a second-order polynomial fitting of $1/M$, we find that the extrapolated $\left | P_{bb}(r) \right |$ shows a power-law decay with distance  $\left | P_{bb}(r) \right |\sim r^{-K_{SC}}$, with the Luttinger exponent $K_{SC}\approx 0.76$. Similar results are obtained for correlations with other bonds, and also for different $L_{x}$ or  $L_{y}=4$  systems (see Supplemental Sec. III. A~\cite{SuppMaterial}).  $K_{SC}\lesssim 1$ holds for SC2 phase, indicating a strong divergent SC susceptibility in the zero-temperature limit~\cite{jiang2021superconductivity}.

We then turn to the density-density  $D(\mathbf{r})=<\widehat{n}_{\mathbf{r}_{0}} \widehat{n}_{\mathbf{r}_{0}+ \mathbf{r}}>-<\widehat{n}_{\mathbf{r}_{0}}><\widehat{n}_{\mathbf{r}_{0}+ \mathbf{r}} >$ and single-particle $G(\mathbf{r})=\sum _{\sigma }< \widehat{c}^{\dagger }_{\mathbf{r}_{0},\sigma } \widehat{c}_{\mathbf{r}_{0}+\mathbf{r},\sigma }>$ correlations. As shown in Fig.~\ref{Fig_SC_correlation}(b), the $\left |D(r) \right |$ decays with a power-law relation at long distance  using extrapolated data. The Luttinger exponent for density-density correlations is $K_{CDW}\approx 1.50$ much larger than $K_{SC}$. 
Both   $\left | P_{bb}(r) \right |$  and  $\left |D(r) \right |$ show similar spatial oscillations consistent with the  electron  density  oscillation  in  real  space (see Supplemental Sec. V~\cite{SuppMaterial}).
Similarly, the single-particle Green function $\left |G(r)\right |$ can also be fit into power-law behavior [Fig.~\ref{Fig_SC_correlation}(b)].
Interestingly, we identify a crossover for $\left |G(r)\right |$ with the increase of $J_{2}$. At smaller $J_{2}=0.05$ in the SC2 regime, we observe an exponential decay in $\left |G(r)\right |$ with a short correlation length $\xi_G\approx 2.41$ which is consistent with the gapped isotropic TSC state as shown in Fig.~\ref{Fig_SC_correlation}(c). With the increase of $J_{2}$, the correlation length for $\left |G(r)\right |$ increases to $\xi_G\approx 8.36$ larger than $L_{y}$, which could also be fitted by a power-law decay [as shown in Fig.~\ref{Fig_SC_correlation}(b)] at $J_{2}=0.1$ for fixed  $J_{\chi}=0.05$ on different systems $N=24\times6$ and $36\times 6$.
The evolution of the single-particle correlation length is a signature of the evolution of the quasiparticle excitation gap, which gradually reduces
with the increase of the $J_{2}$ approaching the quantum phase transition from gapped SC state to a nodal nematic SC state as we will address further in Sec.~\ref{symmetry_evolutions}.
In comparison, the spin-spin correlations remain exponentially decay with a short correlation length $\xi _{S}\approx 1.53$  ($2.06$) as shown in the main panel (inset) in Fig.~\ref{Fig_SC_correlation}(d) at $J_{2}=0.1$ ($0.05$), indicating a finite spin gap which protects the SC state. 
These results provide compelling evidence for the robust SC pairing correlations as dominant correlations for SC phases
with $C=2$, which are  the quasi-1D descendent states of  2D topological superconductors.

Now we  discuss the  features of various correlations at small $J_{2}$, where the SC1 phase is identified with the Chern number $C=1$.
The SC pairing correlations are shown in Fig.~\ref{Fig_SC1}(a), where the magnitude of the SC pairing correlations  $\left | P_{bb}(r) \right |$ show power-law behavior with the Luttinger exponents $K_{SC}\approx 1.09$ and $0.59$ for $J_{2}=0$ and $0.01$, respectively, obtained with a fixed $J_{\chi}=0.05$ for $N=48\times 4$ system.
In comparison,  as shown in Fig.~\ref{Fig_SC1}(b), the $\left |D(r) \right |$ decays with a power-law relation with larger exponents ($K_{CDW}\approx 1.56$ and $1.39$),
while the  $\left |G(r)\right |$ (Fig.~\ref{Fig_SC1}(c)) and  $|S(r)|$ (Fig.~\ref{Fig_SC1}(d)) both decay exponentially with small correlation lengths.
These results indicate that SC1 phase has dominant SC correlations observed for $L_{y}=4$ cylinders.
We also confirm  power-law SC correlations for a wider system of $N=20\times 6$ at $J_{2}=0.01, J_{\chi }=0.05$ with good numerical convergence for this possible gapless phase.  
Interestingly, there are stronger spin correlations for the $L_{y}=6$ system indicating a vanishing or very small spin gap
(see Supplemental Sec. III. D~\cite{SuppMaterial}), which is consistent with the fact that the SC1 phase has a spin gap-closing transition from $C=2$ phase.

We further confirm that the SC0 phase has robust quasi-long-range SC pairing correlations with a Luttinger exponent $K_{SC}\approx 1.10$ at $J_{2}=J_{\chi}=0.2$ dominating the density-density correlations. This phase can be smoothly connected to the d-wave phase identified by doping the $J_{1}$-$J_{2}$ model~\cite{jiang2021superconductivity} with lager $J_{2}$ (see Supplemental Sec. III. C~\cite{SuppMaterial} for more details).

There are other competing quantum phases and additional quantum phase transitions 
as we reduce the three spin chiral interactions to zero in the small $J_{2}$ regime. For example, at $J_{2}=J_{\chi }=0$, 
the ground state is dominated by charge stripe and spin fluctuations with suppressed pairing correlations.
Furthermore,  additional results for $L_{y}=4$ with small $J_{2}$ for a possible pairing density wave SC phase~\cite{peng2021gapless}, and a d-wave SC phase co-existing with phase separation are shown in the Supplemental Sec. III. F~\cite{SuppMaterial}.
These results provide further support that a small but finite chiral interaction plays an important role in stabilizing the topological SC phases for the systems we have studied.

\subsection{Pairing symmetry for topological and nematic SC phases}
\label{pairing_symmetry}
\begin{figure}
\centering
\includegraphics[width=1\linewidth]{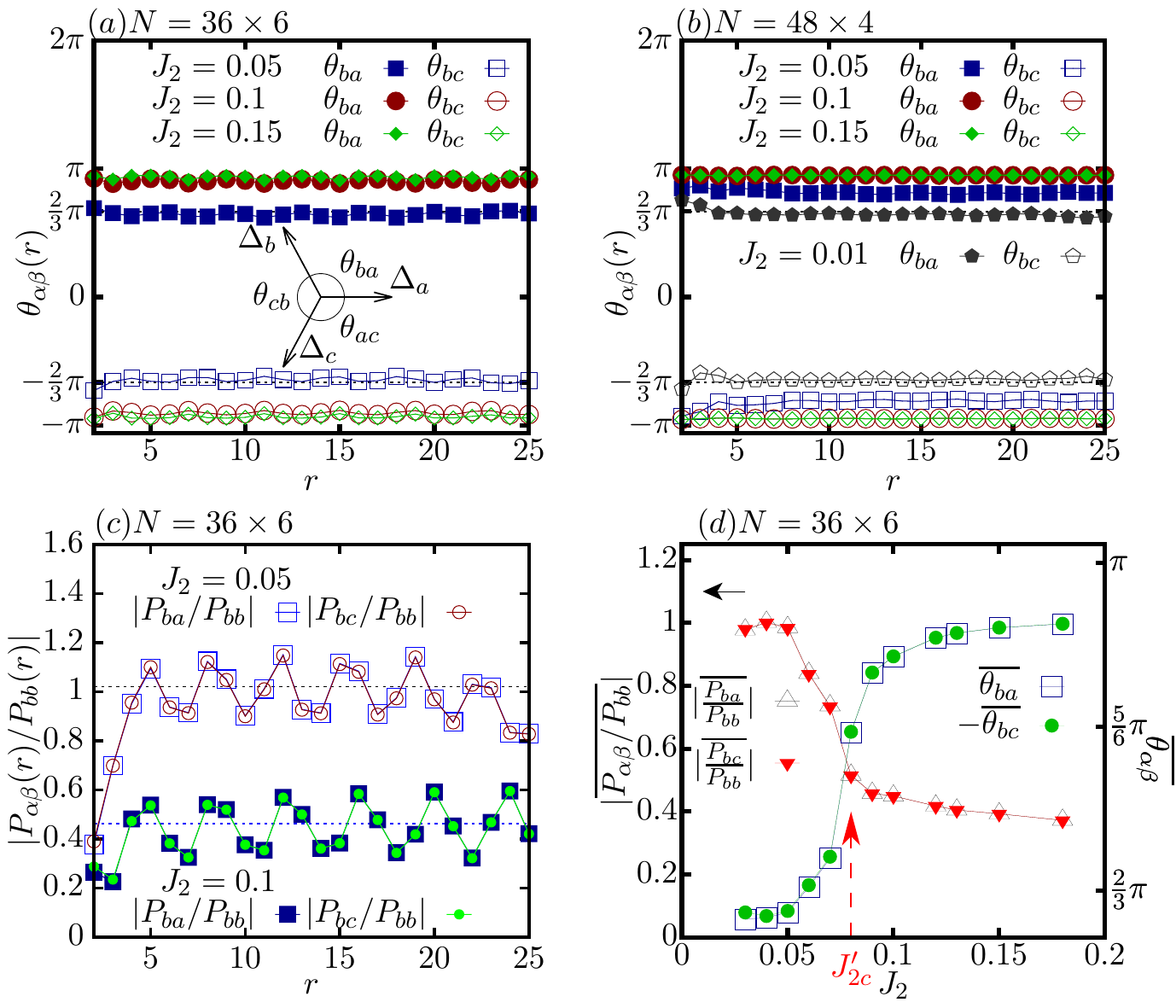}
\caption{Transition from the $d+id$-wave SC2 phases to SC0 phase. (a) The spatial dependence of relative phases for SC  order parameters for various $J_{2}$ on the $L_{y}=6$ ($L_{x}=36$) system. $\mathbf{r}$ is chosen along the $x$ direction $\mathbf{r}=(r,0)$.  (b) The relative phases for SC order parameters on the $L_{y}=4$ ($L_{x}=48$) system.  (c) The ratio of the magnitudes of different SC  correlations. The dashed line indicates the average over distances. (d) The spatial average values of the relative phases of SC order parameters and the ratios of  magnitudes of different SC correlations for various $J_{2}$. All results are obtained at $J_{\chi }=0.05$ with bond dimension $M=10000$ ($M=8000$) on $L_{y}=6$ ($L_{y}=4$) systems. The doping level is $\delta=1/12$.}
\label{Fig_phase_transition}
\end{figure}

The SC pairing symmetry can be identified by the phases of the pairing correlations for different bonds.
To extract the phase we rewrite the SC pairing correlation as $P_{\alpha \beta }(\mathbf{r})=\left | P_{\alpha \beta }(\mathbf{r}) \right |e^{i\phi _{\alpha \beta }(\mathbf{r})}$ where $\phi _{\alpha \beta }(\mathbf{r})$ is the phase for the correlation, and  the pairing order parameter as $\Delta_{\alpha}(\mathbf{r})=\left | \Delta_{\alpha}(\mathbf{r}) \right |e^{i\theta _{\alpha}(\mathbf{r})}$.  Using the definition of the pairing correlations, we obtain $\theta_{\alpha\beta}(\mathbf{r})=\theta_{\alpha}(\mathbf{r})-\theta_{\beta}(\mathbf{r})=\phi_{\alpha\alpha}(\mathbf{r})-\phi_{\alpha\beta}(\mathbf{r})$
as the relative phases of the pairing order parameters  for different nearest-neighbor bonds; see illustrations in the inset in Fig.~\ref{Fig_phase_transition}(a). The  pairing symmetry for  different SC states is illustrated in Fig.~\ref{Fig_phase_transition}(a) on $N=36\times6$ systems. At $J_{2}=J_{\chi}=0.05$ in the SC2 regime,  $\theta _{\alpha \beta }(r)$ remains almost independent of $r$ and the phases for order parameters are nearly quantized to $[{\theta }_{bb },{\theta }_{bc},{\theta }_{ba}]\approx [0,-0.64\pi, 0.65\pi] \approx [0,-\frac{2}{3}\pi,\frac{2}{3}\pi]$, which represents an isotropic $d+id$ wave  with $C_3$ rotational symmetry. The angles ${\theta }_{ba}$ and ${\theta }_{bc}$ are closer to $\pm 2\pi/3$ on wider system $L_{y}=6$  compared with $L_{y}=4$ results (with $L_{x}=48$) as shown in  Figs.~\ref{Fig_phase_transition}(a) and \ref{Fig_phase_transition}(b). At larger $J_{2}=0.15$ with $J_{\chi}=0.05$ in the SC0 phase, these phases become $[{\theta }_{bb },{\theta }_{bc },{\theta }_{ba }]\approx [0,-0.93\pi, 0.93\pi]\approx [0,-\pi, \pi]$, which are nearly independent of  $L_{y}=4$ or $6$, suggesting  a $d$-wave SC state consistent with the Chern number $C=0$ for this phase.

On the other hand,  the relative strength of the SC correlations for different bonds also evolves with the increase of $J_{2}$. As shown in Fig.~\ref{Fig_phase_transition}(c), $|P_{ba}(r)/P_{bb}(r)|$ and $|P_{bc}(r)/P_{bb}(r)|$ have spatial oscillations, and  remain almost a constant average as $r$ increases, which suggests a power-law-decaying behavior of $\left | P_{ba}(r) \right |$ and $\left | P_{bc}(r) \right |$ with the same exponents  $K_{SC}$. At smaller $J_{2}=0.05$, the ratios $|P_{ba}(r)/P_{bb}(r)|$ and $|P_{bc}(r)/P_{bb}(r)|$ are close to $1.0$, while they drop to around $0.46$  observed at $J_{2}=0.1$, keeping the same $J_{\chi}=0.05$.
 We further show an example of the SC1 phase at  $J_{2}=0.01$  and $J_{\chi}=0.05$ on $L_{y}=4$ in Fig.~\ref{Fig_phase_transition}(b) with nearly quantized phases $[{\theta }_{bb },{\theta }_{bc},{\theta }_{ba}]\approx [0,-0.64\pi, 0.64\pi]\approx [0,-\frac{2}{3}\pi,\frac{2}{3}\pi]$.   From mean-field theory, the isotropic SC1 state with $C=1$ has nodal quasiparticle excitations~\cite{zhou2008nodal} and   we leave the full nature of this state to future study due to the increased  difficulty of converging the SC correlations for this critical phase on $L_{y}=6$.

\section{Symmetry evolution and phase transitions}
\label{symmetry_evolutions}

 \subsection{Evolution of the pairing order parameters}
 \label{pairing_order_evolution}
 
Now we focus on the symmetry evolution of the pairing order parameters from SC2 phases to SC0 phase.
 As shown in Fig.~\ref{Fig_phase_transition}(d), the $-\overline{\theta _{bc }}$ and $\overline{\theta _{ba}}$
 averaged over the middle $24$ columns of the system with $L_{x}=36$
 increase monotonically from $\frac{2}{3}\pi $ towards $\pi$ as $J_{2}$ increases.
At the same time,  we find direct evidence of the increased nematicity as $J_{2}$  increases, which is identified by the ratio of the 
 magnitudes of SC pairing correlations  for different bonds. As shown in Fig.~\ref{Fig_phase_transition}(d), the spatial averaged ratios  $\overline{ \left |P_{ba} / P_{bb} \right| }$ and $ \overline{ \left |{P}_{bc} / {P}_{bb} \right |}$ decrease monotonically from 1 to around 0.4 at larger $J_{2}$  side.
 A transition from an isotropic TSC phase to a nematic SC phase takes place inside the SC2 regime as indicated by the arrow in  Fig.~\ref{Fig_phase_transition}(d) pointing to a critical $J_{2c}'$, where both the 
$\overline{ \left |P_{ba} / P_{bb} \right| }$ and $ \overline{ \left |{P}_{bc} / {P}_{bb} \right |}$
decrease quickly to the near saturated value. This feature revealed by the nematicity evolution shows a transition inside the SC2 regime, indicating that a nematic TSC phase emerges for $J_{2c}' < J_{2} <J_{2c}^{(2)}$.
As shown in Fig.~\ref{Fig_phase_diagram}(b), we identify a finite regime with increased nematicity in the SC pairing correlations while its topological nature remains the same with the Chern number $C=2$, which is consistent with  a nematic TSC state emerging within $C=2$ class of SC phases. The phase boundary is determined by the quick increase of $\overline{\theta _{ba}}$ to around $\frac{5}{6}\pi$. The emergent nematic TSC state is an analog state to the nematic fractional quantum Hall state with gapless collective excitations~\cite{haldane2011geometrical,you2014theory,yang2017anisotropic,regnault2017evidence}. With further increase of $J_{2}$, the topological quantum phase transition takes place where the nematic $d$-wave SC state is recovered in  the SC0 phase.

\subsection{Nature of quantum phase transitions}
\label{phase_transition}
 
\begin{figure}
\centering
\includegraphics[width=1\linewidth]{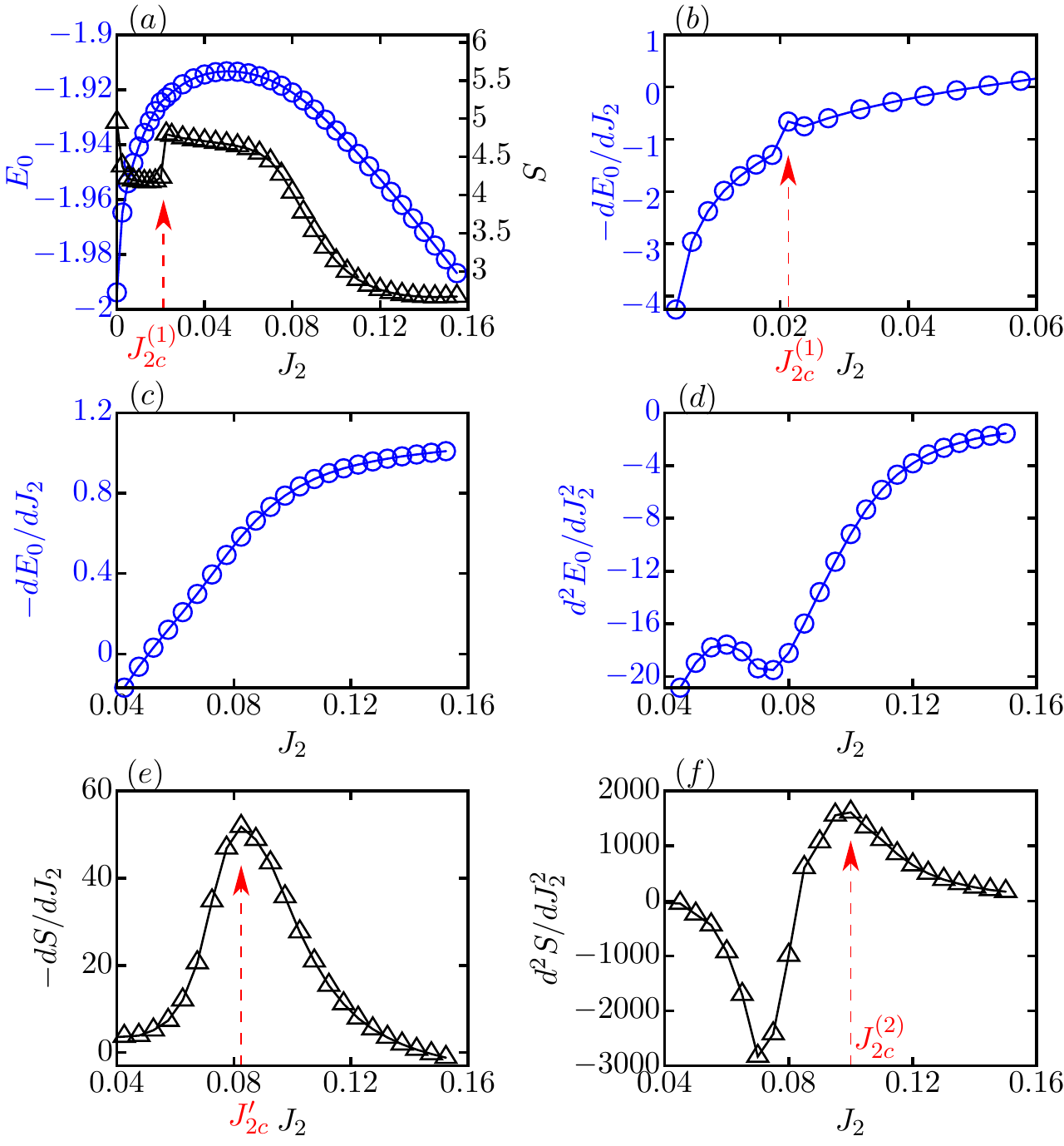}
\caption{The energy per site $E_{0}$ and entanglement entropy $S$ for various $J_{2}$ obtained at fixed $J_{\chi }=0.05$ on a $N=16\times 6$ cylinder. (a) The energy and entanglement entropy, where $J_{2c}^{(1)}$ is identified from  the small jump in $S$. 
(b) The first order derivative of $E_{0}$ with respect to $J_{2}$ where $J_{2c}^{(1)}$ is  identified at the discontinuity point. (c) The first order derivative of $E_{0}$ for larger $J_{2}$. (d) The second order derivative of $E_{0}$ for larger $J_{2}$. (e) The first order derivative of $S$ where $J_{2c}'$ is identified as the peak. (f) The second order derivative of $S$ where $J_{2c}^{(2)}$ is identified as the peak. The results are obtained with $M=10000$. The doping level is $\delta=1/12$.}
\label{Fig_nature_transiton}
\end{figure}

We now explore the nature of quantum phase transitions by following the energy and entanglement entropy evolution along the parameter line of $J_{\chi}=0.05$. To  calculate the entanglement entropy $S$, the cylinder is cut into two equal halves and $S$ is  obtained  from the eigenvalues $\lambda_i$ of the reduced density matrix $S =-\sum _{i} \lambda_{i}log (\lambda _{i})$.
As shown in Fig.~\ref{Fig_nature_transiton}(a), the energy per site $E_0$ shows 
a small kink and the entropy $S$ shows a large jump at $J_{2c}^{(1)}\approx 0.021$ which is very close to the transition point between $C=1$ and $C=2$ phase, indicating a first order transition between SC1 and the isotropic TSC (SC2) phase. The first order transition is further revealed by $-dE_{0}/dJ_{2}$ given in Fig.~\ref{Fig_nature_transiton}(b), which shows discontinuity around $J_{2c}^{(1)}$. Two other transitions from the isotropic TSC to nematic TSC and from nematic TSC to d-wave SC are continuous transitions with smooth evolution of $E_{0}$ and $S$ (Fig.~\ref{Fig_nature_transiton}(a)), and their derivatives (Figs.~\ref{Fig_nature_transiton}(c) - (f)). Interestingly, the transition point between the isotropic TSC to nematic TSC is indicated by the peak in $-dS/dJ_{2}$, which is very close to the one identified by the nematicity in SC pairing correlations. The transition between nematic TSC and the nematic d-wave SC may be identified by the peak in $d^{2}S/dJ_{2}^{2}$, which is shifted from the one identified by the Chern number with flux insertion into a very long cylinder studied in the infinite DMRG.  This may be explained by
the finite size effect  for identifying higher order transitions where the peaks in entropy  usually shift with system lengths~\cite{zhong2009quantum,zhang2021fidelity}.
 
\subsection{Evolution of the spin correlations}
 \label{spin_correlations}
\begin{figure}
\centering
\includegraphics[width=1\linewidth]{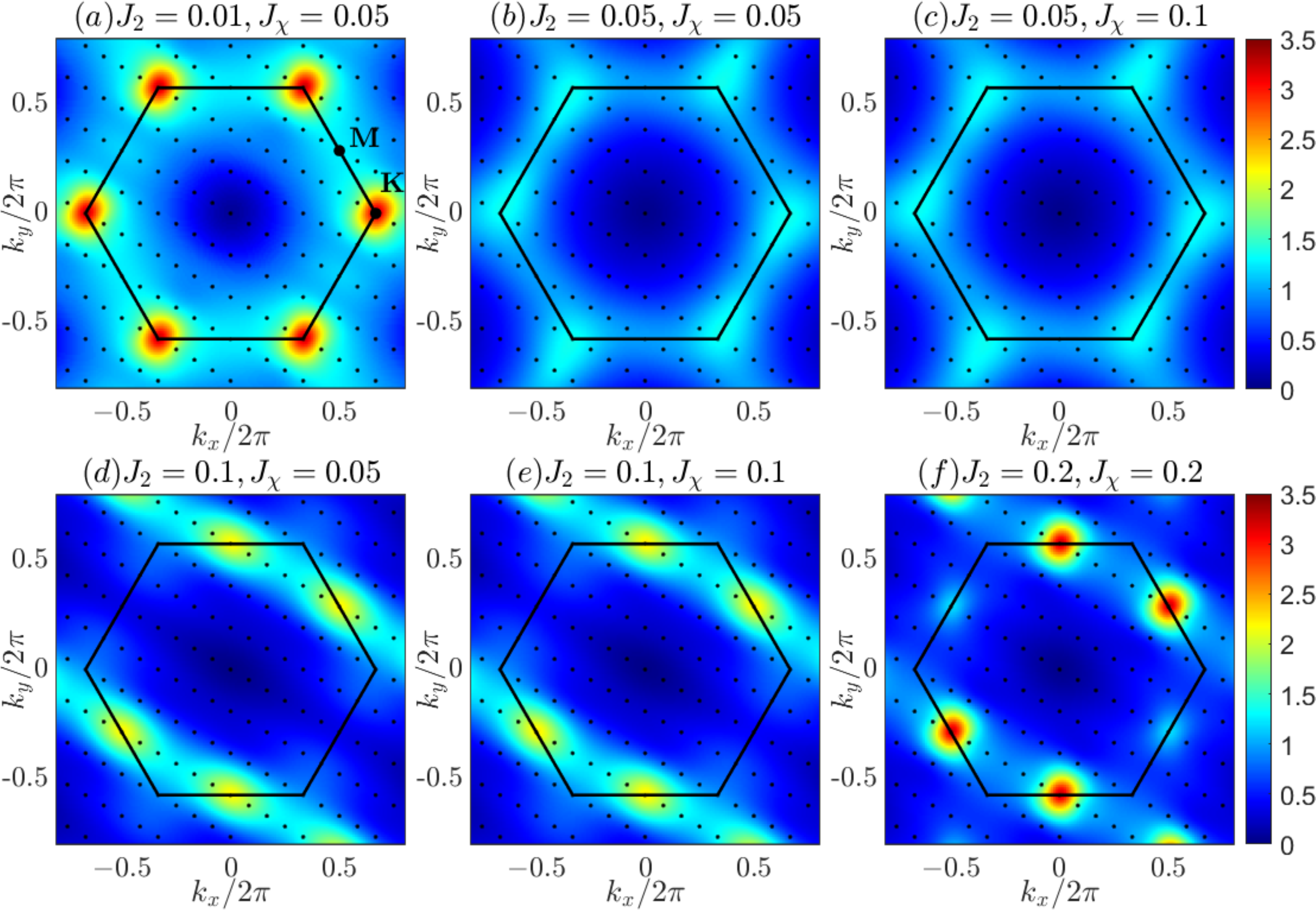}
\caption{The spin structure factor $S(\mathbf{k})$ obtained on an $N=36\times 6$ cylinder using correlations from middle $N'=12\times 6$ sites.  (a) SC1 phase at $J_{2}=0.01$. (b), (c) Isotropic $d+id$-wave TSC phase at $J_{2}=0.05$ with near-isotropic structure. (d), (e) Nematic TSC phase at $J_{2}=0.1$.  (f) SC0 state at $J_{2}=0.2$.  The first Brillouin zone is indicated by the solid line with the $\mathbf{M}$ and $\mathbf{K}$ points marked, and the black dots represent the allowed discrete momenta for the finite system with $12\times 6$ sites. The results are obtained with $M=10000$. The doping level is $\delta=1/12$.}
\label{Fig_spin_structure}
\end{figure}

The evolution of spin correlations can be studied through the spin structure factor defined as   $S(\mathbf{k})$ $=\frac{1}{N^{'}}\sum_{i,j}\left \langle \mathbf{S}_{i} \cdot \mathbf{S}_{j} \right \rangle e^{i\mathbf{k}\cdot (\mathbf{r}_{i}-\mathbf{r}_{j})}$ for a system with $N=36\times 6$, where $i$ and $j$ are summed over the middle $N^{'}=2L_{y} \times L_{y}$ sites to avoid boundary effects. As shown in Fig.~\ref{Fig_spin_structure}(a),  $S(\mathbf{k})$ has strong peaks at the $\mathbf{K}$ points representing strong $120^{\circ}$ AFM fluctuations at short distances for $J_{2}=0.01,J_{\chi}=0.05$ inside the SC1 phase, while the spin correlations exponentially decay at long distance for all three classes of SC phases (see Supplemental Sec. III. E~\cite{SuppMaterial} for details). With a larger $J_{2}=0.05$ in the SC2 phase, $S(\mathbf{k})$ becomes nearly featureless with some intensity around the Brillouin zone boundaries as shown in Figs.~\ref{Fig_spin_structure}(b) and \ref{Fig_spin_structure}(c), which is consistent with an isotropic $d+id$-wave TSC state. As $J_{2}$ further increases to $0.1$, moderate peaks appear in $S(\mathbf{k})$ at the $\mathbf{M}$ points with nematicity as seen in Figs.~\ref{Fig_spin_structure}(d) and \ref{Fig_spin_structure}(e), where the SC pairing order parameters also become anisotropic [Fig.~\ref{Fig_phase_transition}(d)]. Further increasing $J_{2}$ into the SC0 phase, the spin fluctuations appear as brighter  peaks in $S(\mathbf{k})$ at the $\mathbf{M}$ points, with stronger stripe fluctuations  as shown in Fig.~\ref{Fig_spin_structure}(f). The emerging picture is that the spin nematicity tuned by hole dynamics (see more details in the Supplemental Sec. IV~\cite{SuppMaterial}) and spin interactions with the increase of $J_{2}$ (and $t_{2}$) are the determining forces in driving the quantum phase transitions between different SC phases.

\section{Summary and discussion}
\label{summary}
We have extensively studied the ground state of the lightly doped extended $t$-$J$-$J_{\chi}$ model on the triangular  lattice based on the state-of-the-art DMRG method and   unraveled  a global picture of emergent unconventional superconductivity in systems with chiral interactions that could be induced by an external magnetic field~\cite{jiang2020topological}. We identify three classes of superconducting phases (SC1, SC2, and SC0)  characterized by different topological Chern numbers and pairing symmetries. As next nearest neighbor hopping $t_{2}$ and the related spin coupling $J_{2}$ increase, the critical SC1 state  with  Chern number $C=1$ has a transition to the isotropic TSC
phase, which is a gapped topological $d+id$-wave superconductor with $C=2$. 
 With further increase of $t_{2}$ and $J_{2}$, the  isotropic TSC state has a transition to a nematic TSC state in the smaller $J_{\chi}$ regime, which is an analogy of the nematic fractional quantum Hall state with broken rotational symmetry~\cite{you2014theory,yang2017anisotropic,regnault2017evidence}. A topological phase transition from the $C=2$ TSC states to the nematic d-wave SC0 state with  $C=0$ occurs for larger  $t_{2}$ and $J_{2}$. The hole dynamics tuned by next-nearest-neighboring hoppings and spin couplings drives the topological quantum phase transitions between different SC phases, and a small chiral interaction $J_{\chi}\approx 0.01$ stabilizes the SC states.

The TSC is a long-sought state, and  it was conjectured that such a SC state may be realized  by doping a CSL~\cite{kalmeyer1987equivalence, wen1989chiral} if it can win over other competing states varying from a chiral metal~\cite{song2021doping, zhu2022doped} to a fractionalized Wigner crystal~\cite{jiang2017holon,peng2021}. Despite intensive efforts in searching for such  a TSC state in strongly correlated systems  during the past decades, there is only one established example  by unbiased numerical studies  of the  $t$-$J$-$J_{\chi}$ model~\cite{jiang2020topological}. The identified TSC state also showed some  instability on a wider cylinder ($L_{y}=6$)~\cite{jiang2020topological} before adding next-nearest-neighboring hopping, indicating that it is near a phase boundary. In this work, we uncover a global phase diagram  for  the same  model and unravel  two distinct classes of TSC phases with Chern numbers $C=1,2$, and a nematic d-wave SC phase with $C=0$. The new insight to the mechanism of the doping-induced TSC is that  the TSC can emerge by doping a correlated Mott insulating state with $120^{\circ}$ AFM besides doping a CSL  state. Importantly, the hole dynamics changes the spin background and induces a topological quantum phase transition upon doping a magnetic ordered state, widening the opportunity for discovering TSC in triangular compounds. On the other hand, nematic SC with $C=0$ can emerge from either doping the CSL or magnetic ordered states~\cite{gong2017global}, suggesting  the rich interplay between unconventional SC and spin background.

A recent analytical study~\cite{schrade2021nematic} has identified  topological and nematic SC  in the moir$\acute{e}$ superlattice of twisted bilayer TMD that realizes an effective triangular lattice  model with repulsive interactions. 
While the study addresses the physics in the weak coupling picture with spin-valley fluctuations, it is interesting  to see the  $C=2$ topological $d+id$-wave SC and the nematic $d$-wave SC being discovered, which may  indicate a possible universal picture for SC on triangular lattice with repulsive interactions. It would be exciting to examine the extended Hubbard model with short-range Coulomb interactions for such systems from weak to strong couplings~\cite{arovas2022hubbard}, which can make further predictions for TMD systems. In light of the theoretical prediction of the SU(4) CSL~\cite{zhang20214} in time-reversal-invariant TMD bilayers, we anticipate TSC to be a strong competing state and a rich phase diagram to be revealed. Besides the triangular lattice, it will also be interesting to search for possible TSC states in other systems including  kagome compounds, where 
Mott insulators show similar rich physics  with emergent CSL~\cite{gong2014}.

\begin{acknowledgments}
We thank Z. Q. Wang and Yahui Zhang for stimulating discussions. 
This work was supported by  the U.S. Department of Energy, Office of Basic Energy Sciences under Grant No. DE-FG02-06ER46305. 
\end{acknowledgments}




\bibliography{Tr_tJJchi}

\begin{thebibliography}{75}
\bibitem [{not()}]{note1}%
  \BibitemOpen
  \href@noop {} {}\bibinfo {howpublished} {For $J_2<0.01$ or $J_{\chi}<0.01$,
  it becomes increasingly difficult to obtain converged ground state or
  identify phase boundaries.}\BibitemShut {Stop}%
\end{thebibliography}



\newcommand{\beginsupplement}{%
        \setcounter{table}{0}
        \renewcommand{\thetable}{S\arabic{table}}%
        \setcounter{figure}{0}
        \renewcommand{\thefigure}{S\arabic{figure}}%
        \setcounter{section}{0}
        \renewcommand{\thesection}{\roman{section}}%
        \setcounter{equation}{0}
        \renewcommand{\theequation}{S\arabic{equation}}%
        }
\clearpage
\beginsupplement
\onecolumngrid
\begin{center}
\Large Supplementary for ``Topological chiral and nematic superconductivity  by doping Mott insulators on triangular lattice''
\end{center}


\section{Robust spin pumping  with flux insertion}
\label{SM_chern}

\begin{figure}
\centering
\includegraphics[width=1\linewidth]{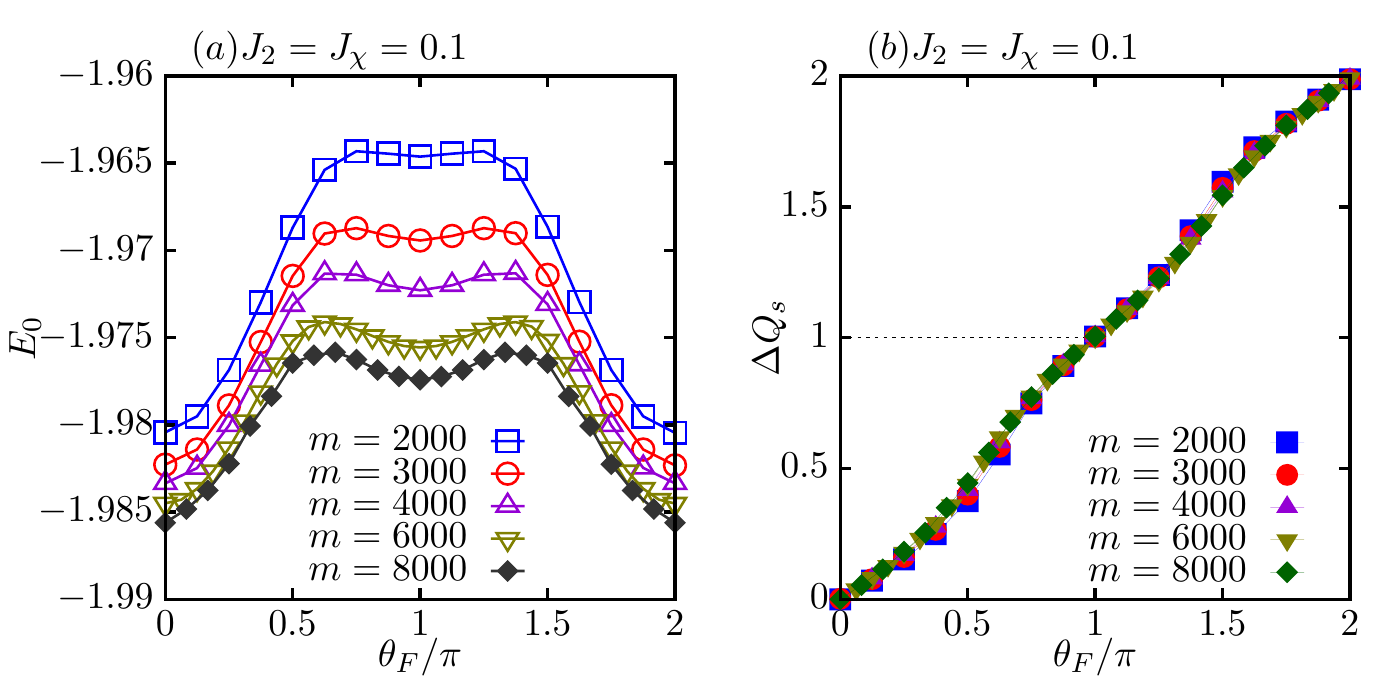}
\caption{(a) The energy per site $E_{0}$. (b) The spin pumping $\Delta Q_{s}$ versus flux $\theta_F$ for various bond dimension $m$. Results are obtained at $J_{2}=J_{\chi }=0.1$ and $\delta = 1/12$ on $L_{y}=6$ infinite  cylinder.}
\label{FigS_chern}
\end{figure}

The flux insertion method has been extensively applied to different fractional quantum Hall  and chiral spin liquid systems to study the topological nature. Here we demonstrate the robustness of the method when we apply it to the topological superconducting states  of the doped triangular $t$-$J$-$J_{\chi}$ model as defined in the main text. Because the inserted flux breaks SU(2) symmetry, we use infinite DMRG with $U(1)\times U(1)$ symmetries for both charge and spin degrees of freedom with a large unit cell commensurate with the doping level. We  check the spin pumping versus inserting flux for various bond dimensions $m$ to confirm the robustness of results. As shown in Fig.~\ref{FigS_chern}(a) at $J_{2}=J_{\chi}=0.1$ and $\delta=1/12$ on $L_{y}=6$ cylinder, the ground state energy per site $E_0$   varies smoothly with  flux $\theta_{F}$, and $E_0(\theta_F)$ curves  have similar shapes for different $m$,  indicating that the flux insertion simulations are less sensitive to bond dimension $m$. Indeed, we find that the pumped spin $\Delta Q_{s}$ evolves with $\theta_{F}$ smoothly with all curves almost fall on top of each other, which is insensitive to the bond dimension $m$. It gives rise to the same quantized Chern number $C=2$ after completing the insertion of one flux quantum for different $m$ as shown in Fig.~\ref{FigS_chern}(b). We use a unit cell of $4\times L_{y}$ sites or $8\times L_{y}$ sites for the doping level $\delta=1/12$ on $L_{y}=6$, which always gives the same quantized results.

\section{Energy convergence  of the ground state with increasing bond dimension}
\label{SM_energy}

\begin{figure}
\centering
\includegraphics[width=0.9\linewidth]{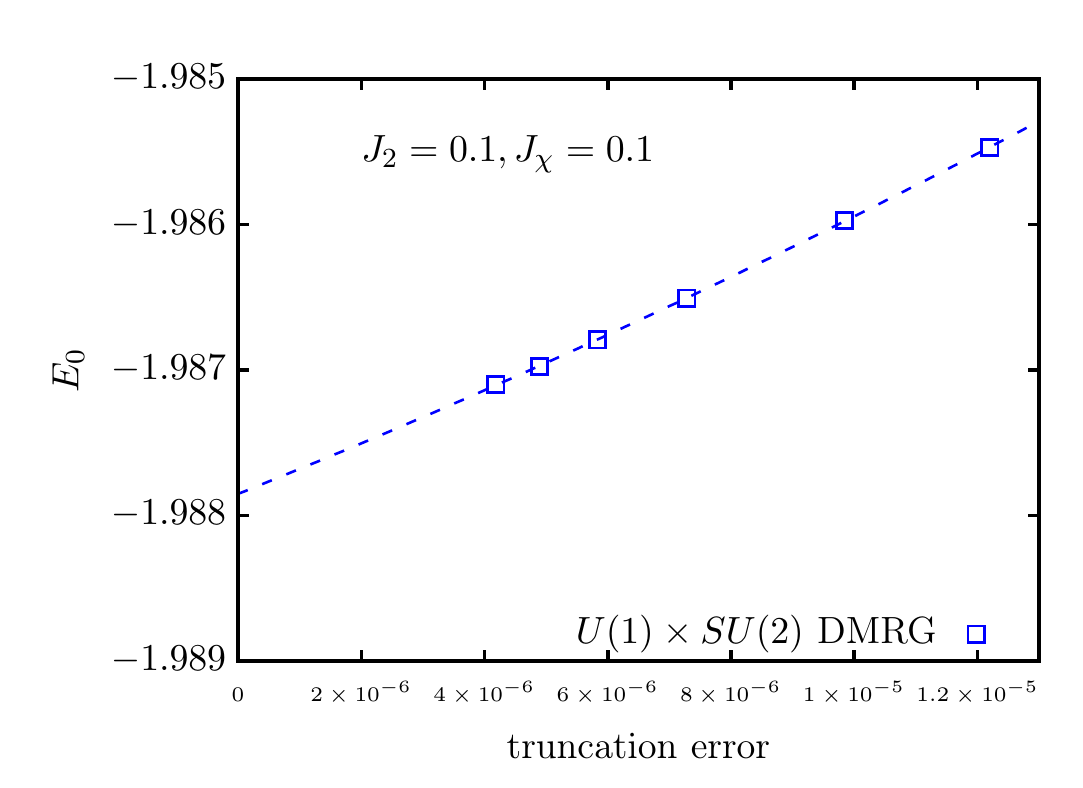}
\caption{The ground state energy per site $E_{0}$ calculated based on  $U(1)\times SU(2)$ DMRG for various bond dimension $M$, obtained at $J_{2}=J_{\chi }=0.1$ and $\delta = 1/12$ for  $N=36\times 6$  system.}
\label{FigS_energy}
\end{figure}

We use the finite size DMRG with the $U(1)\times SU(2)$ symmetries to study the ground state properties of the doped triangular $t$-$J$-$J_{\chi}$ model.
Here we show an example of the ground state energy convergence with the increase of the bond dimensions $M=3000 \sim 12000$ (which are equivalent to 
$m=9000 \sim 36000$ $U(1)$ states) for a system with $N=36\times 6$. As shown in Fig.~\ref{FigS_energy}, we demonstrate the obtained energy per site $E_0$  (measured in the middle $12\times 6$ sites) versus the DMRG truncation error. With $M=12000$, the truncation error is reduced to about  $4\times 10^{-6}$. The extrapolated energy $-1.98785$ is also  close to the lowest energy that we obtain with $M=12000$, indicating the good convergence of our calculations.

\section{More examples of correlation functions in different phases}
\label{SM_correlations_phase}

\subsection{Results of correlation functions in various systems sizes for SC2 phase with $J_{2}=0.1$ and $J_{\chi }=0.05$}

\begin{figure}
\centering
\includegraphics[width=1\linewidth]{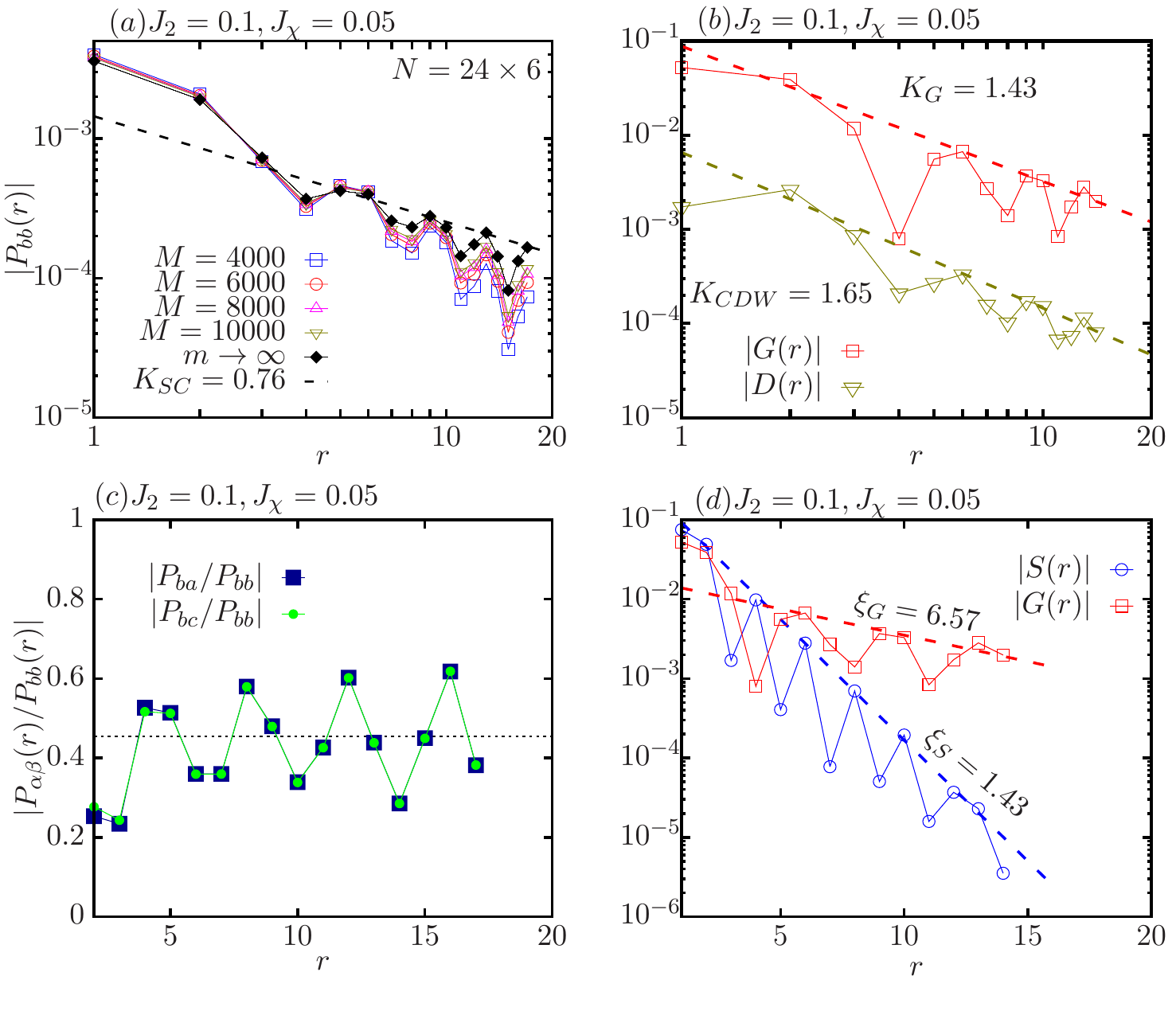}
\caption{Various correlation functions on $N=24\times 6$ system, obtained at $J_{2}=0.1, J_{\chi }=0.05$. $\mathbf{r}$ is chosen along x-direction $\mathbf{r}=(r,0)$. (a) The scaling of SC pairing correlation with bond dimension $M$. The Luttinger exponent is obtained by the straight-line  fitting in the log-log plot  for the extrapolated  infinite $M$ data. (b) The density-density correlation $|D(r)|$ and single particle correlation $|G(r)|$, which are also fit by the power-law relations. (c) The ratio between the   magnitudes of SC pairing correlations for different bonds. The dash line indicates the spatial average over distances.  (d) The exponential fitting of the spin correlation $|S(r)|$ and $|G(r)|$.}
\label{FigS_correlations_J2_01_Jx_005_6_24}
\end{figure}

\begin{figure}
\centering
\includegraphics[width=1\linewidth]{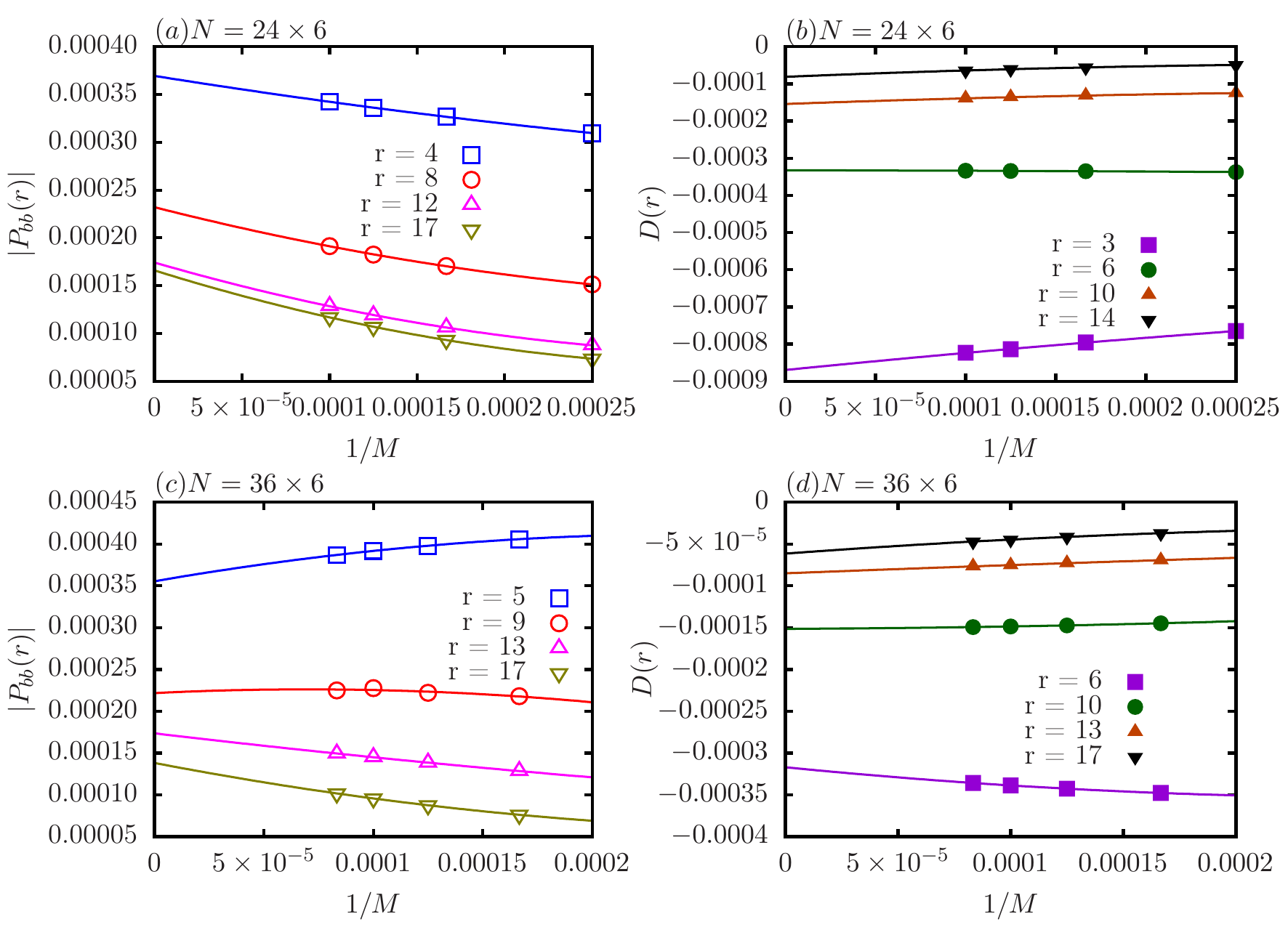}
\caption{The extrapolations of correlation functions at $J_{2}=0.1, J_{\chi }=0.05$ versus $1/M$ for (a) $|P_{bb}|$ and (b) $D(r)$ on $N=24\times 6$ system.
(c) and (d) are similar extrapolations on $N=36\times 6$ system. }
\label{FigS_extrapolation_J2_01_Jx_005_6_24}
\end{figure}

Results on different system sizes are  presented here to compare with the results in the main text at $J_{2}=0.1, J_{\chi }=0.05$ on $N=36\times 6$ system and $\delta = 1/12$. As shown in Fig.~\ref{FigS_correlations_J2_01_Jx_005_6_24}, we study similar correlation functions on a shorter system $N=24\times 6$. The superconductivity (SC) pairing correlations in Fig.~\ref{FigS_correlations_J2_01_Jx_005_6_24}(a) show a quasi-long-range power-law decay after the scaling to $M \rightarrow \infty $, with a Luttinger exponent $K_{SC}\approx 0.76$ the same as the result on $L_{x}=36$ in the main text. The ratio between the magnitudes of SC pairing correlations for different bonds also have almost the same average values as those on $L_{x}=36$, indicating a small finite size effect on the length of $L_{x}$. Furthermore, the density-density correlation $D(\mathbf{r})=<\widehat{n}_{\mathbf{r}_{0}} \widehat{n}_{\mathbf{r}_{0}+ \mathbf{r}}>-<\widehat{n}_{\mathbf{r}_{0}}><\widehat{n}_{\mathbf{r}_{0}+ \mathbf{r}} >$, single particle correlation $G(\mathbf{r})=<\sum _{\sigma }\widehat{c}^{\dagger }_{\mathbf{r}_{0},\sigma }\widehat{c}_{\mathbf{r}_{0}+\mathbf{r},\sigma }>$, and the spin correlation $S(\mathbf{r})=<\widehat{\boldsymbol{S}}_{\mathbf{r}_{0}}\cdot \widehat{\boldsymbol{S}}_{\mathbf{r}_{0}+\mathbf{r}}>$ also show near  the same decaying behaviors as the  results on $L_{x}=36$. Specifically,  we can fit the $|G(r)|$ with both exponential decay and power-law decay  functions, as shown in Figs.~\ref{FigS_correlations_J2_01_Jx_005_6_24}(b) and (d). Under the exponential decay fitting, the correlation length $\xi _{G}\approx 6.57$, which  is already larger than the system  width $L_{y}=6$, which could also be fit with  power-law decay. In comparison, the $|S(r)|$ given in Fig.~\ref{FigS_correlations_J2_01_Jx_005_6_24}(d) has a correlation length of $\xi _{S}=1.43$ which is much smaller than the system width $L_{y}$ indicating fast exponential decay for spin correlations.

\begin{figure}
\centering
\includegraphics[width=1\linewidth]{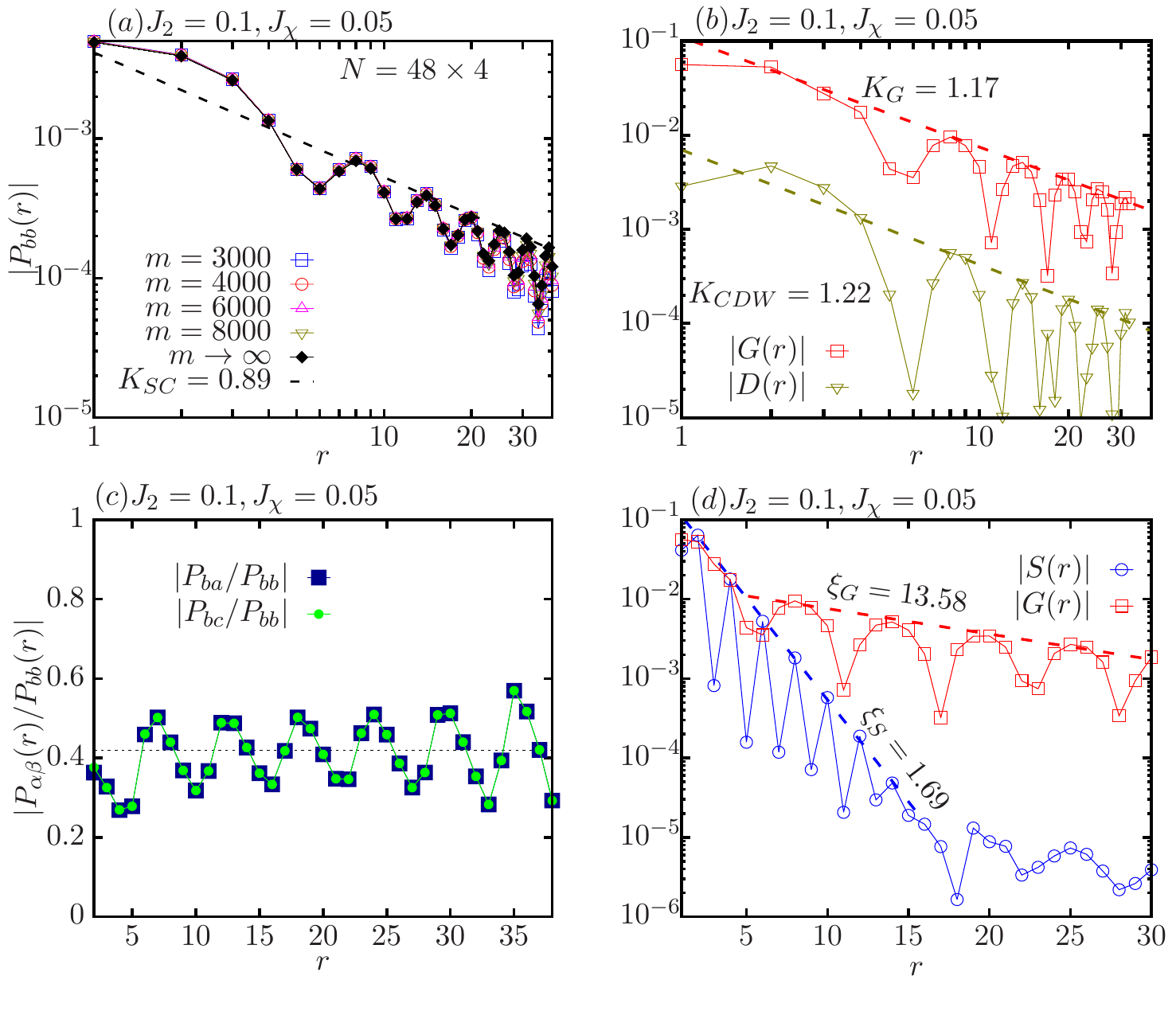}
\caption{Various correlations on $N=48\times 4$ system, obtained at $J_{2}=0.1, J_{\chi }=0.05$. (a) The scaling of SC pairing correlation with bond dimension $M$. The Luttinger exponent is obtained by the straight-line  fitting in the log-log plot for the extrapolated  infinite $M$ data. (b) The density-density correlation $|D(r)|$ and single particle correlation $|G(r)|$, which are also fit by the power-law relations. (c) The ratio between the   magnitudes of SC pairing correlations for different bonds. The dash line indicates the spatial average over distances.  (d) The exponential fitting of the spin correlation $|S(r)|$ and $|G(r)|$.}
\label{FigS_correlations_J2_01_Jx_005_4_48}
\end{figure}

\begin{figure}
\centering
\includegraphics[width=1\linewidth]{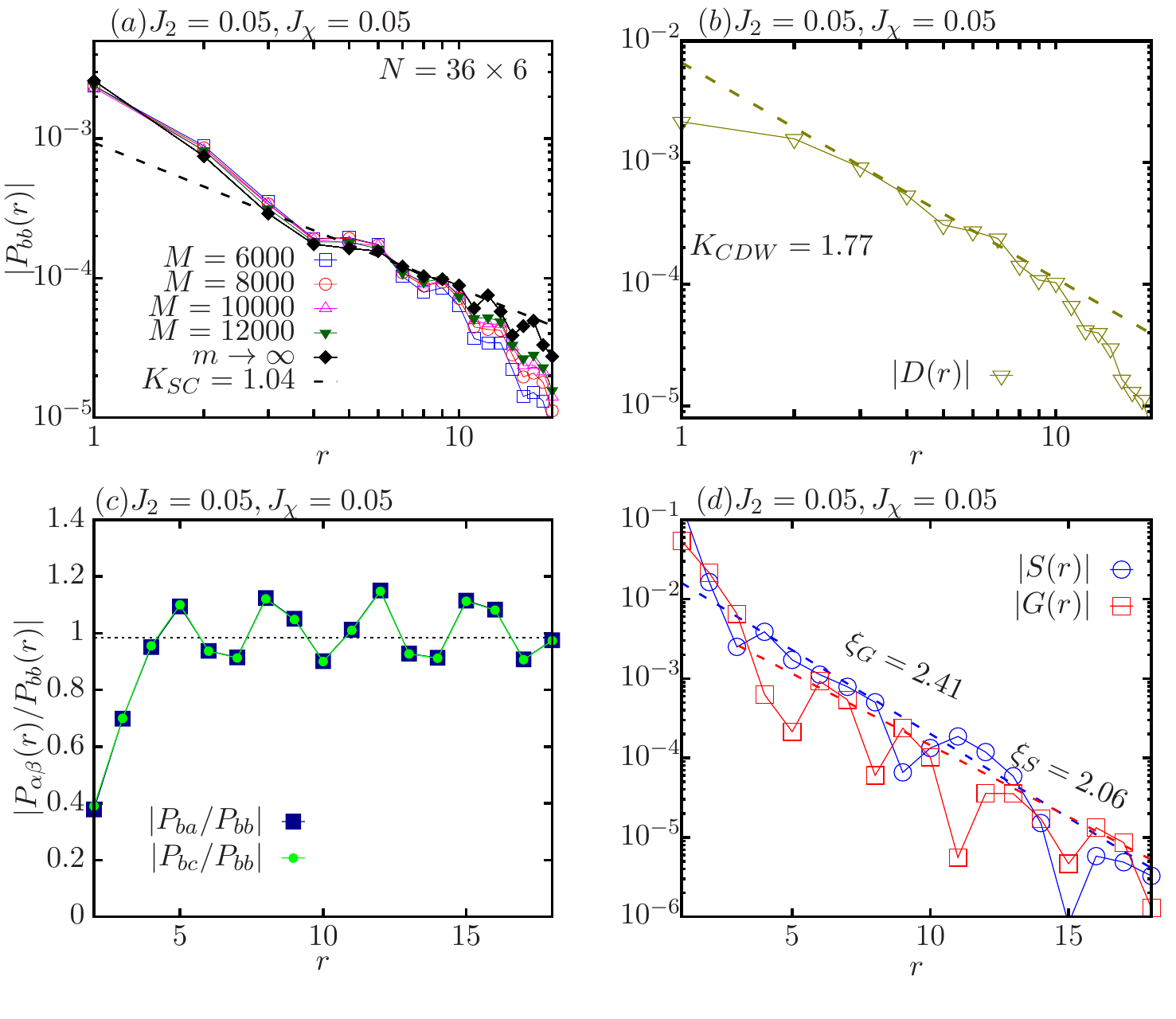}
\caption{Various correlations on $N=36\times 6$ system, obtained at $J_{2}=0.05, J_{\chi }=0.05$. (a) The scaling of SC pairing correlation with bond dimension $M$. The Luttinger exponent is obtained by the straight-line  fitting in the log-log plot for the extrapolated  infinite $M$ data. (b) The density-density correlation $|D(r)|$ is fit by the power-law relation at short distances. The $|D(r)|$ decays faster for $r>10$. (c) The ratio between the   magnitudes of SC pairing correlations for different bonds. The dash line indicates the spatial average over distances.  (d) The exponential fitting of the spin correlation $|S(r)|$ and single particle correlation $|G(r)|$.}
\label{FigS_correlations_J2_005_Jx_005_6_36}
\end{figure}

The finite bond dimension scaling of different physical quantities are used to obtain accurate results of the ground state. Here we show the details in the scaling of correlations versus $1/M$ for two typical examples $|P_{bb}(r)|$ and $D(r)$ using the polynomial fitting up to the second order of $1/M$. As shown in Fig.~\ref{FigS_extrapolation_J2_01_Jx_005_6_24} on different $L_{x}=24$ and $36$ with $L_{y}=6$, the $|P_{bb}(r)|$ and $D(r)$ vary smoothly  with increasing $M$ and can be reliably  extrapolated to the infinite $M$ limit.

Besides studies on $L_{y}=6$, we also find consistent results on $L_{y}=4$ systems. As shown in Fig.~\ref{FigS_correlations_J2_01_Jx_005_4_48}, 
the same decay patterns are found  for all correlation functions comparing $L_{y}=4$ with $L_{y}=6$ results (shown in the main text), indicating a robust quasi-long-range SC order dominating other correlations. In particular, the SC Luttinger exponent $K_{SC}=0.89$ on $L_{y}=4$, which is larger than $K_{SC}=0.76$ on $L_{y}=6$ ($L_{x}=36$) at the same parameter point $J_{2}=0.1, J_{\chi }=0.05$, which suggests a stronger SC on wider cylinders.

\begin{figure}
\centering
\includegraphics[width=1\linewidth]{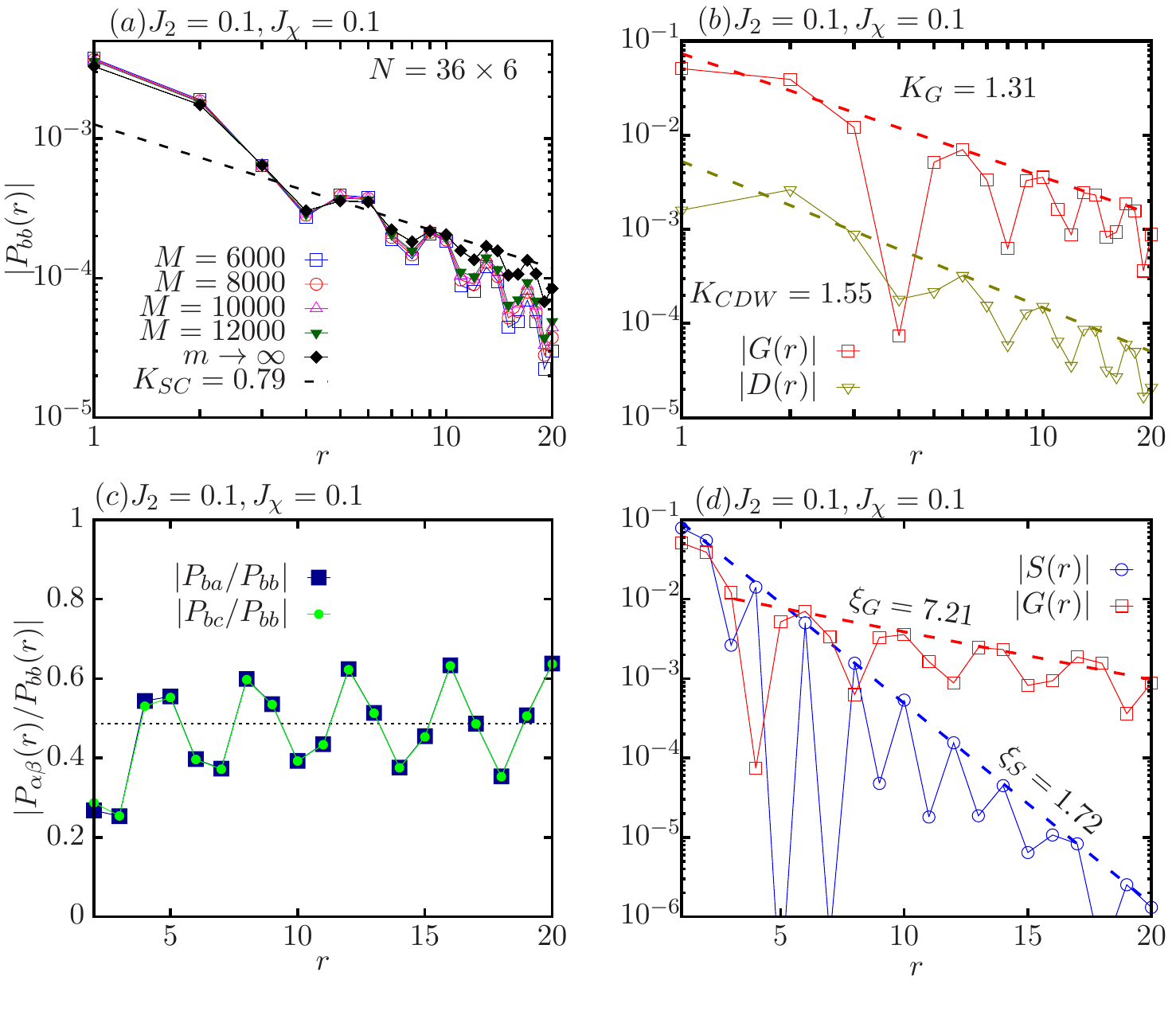}
\caption{Various correlations for $N=36\times 6$ system, obtained at $J_{2}=0.1, J_{\chi }=0.1$. (a) The scaling of SC pairing correlation with bond dimension $M$. The Luttinger exponent is obtained by the straight-line  fitting in the log-log plot for the extrapolated  infinite $M$ data. (b) The density-density correlation $|D(r)|$ and single particle correlation $|G(r)|$, which are also fit by the power-law relations. (c) The ratio between the   magnitudes of SC pairing correlations for different bonds. The dash line indicates the spatial average over distances.  (d) The exponential fitting of the spin correlation $|S(r)|$ and $|G(r)|$.}
\label{FigS_correlations_J2_01_Jx_01_6_36}
\end{figure}

\begin{figure}
\centering
\includegraphics[width=1\linewidth]{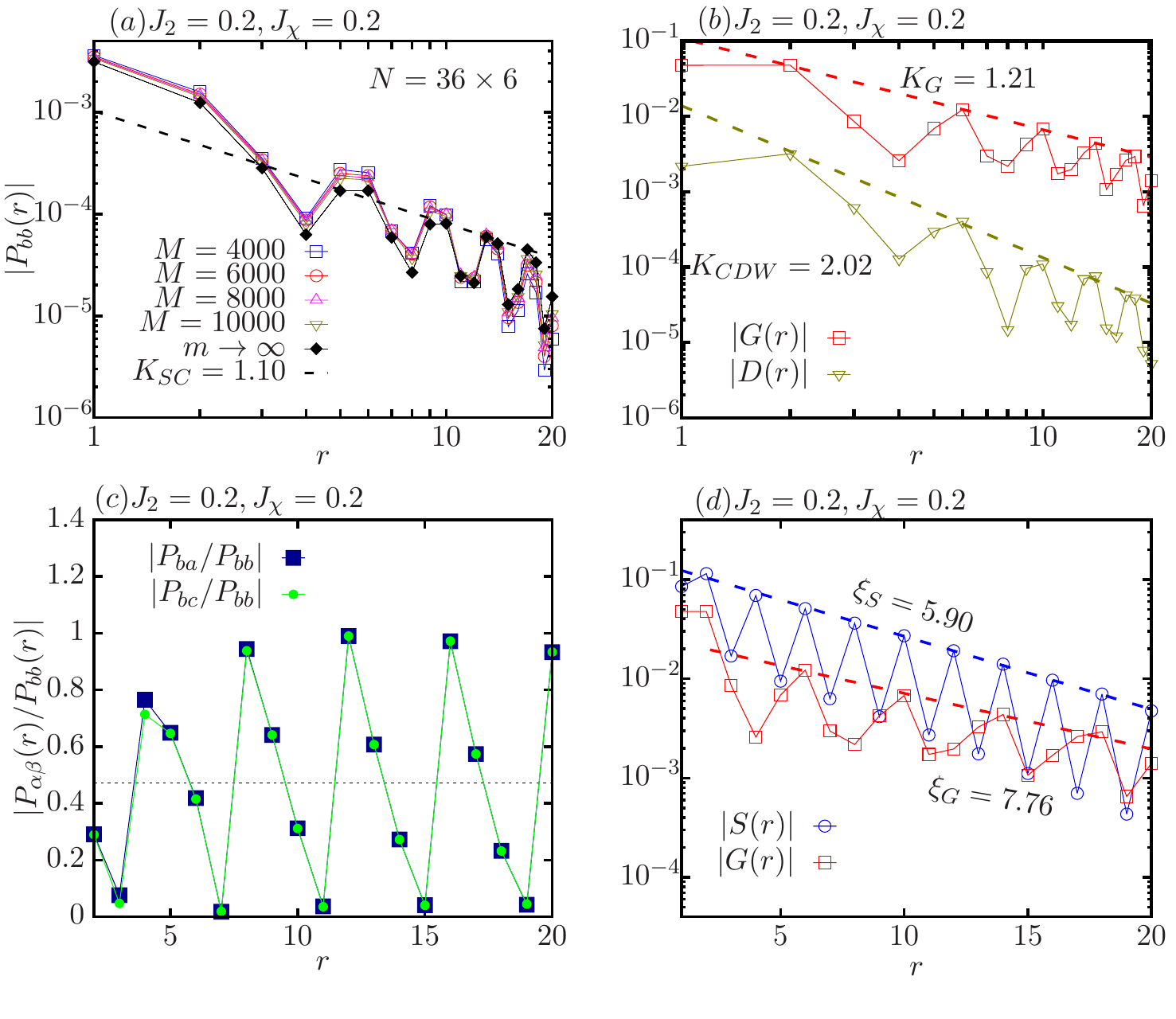}
\caption{Various correlations on $N=36\times 6$ system, obtained at $J_{2}=0.2, J_{\chi }=0.2$. 
(a) The scaling of SC pairing correlation with bond dimension $M$. The Luttinger exponent is obtained by the straight-line  fitting in the log-log plot for the extrapolated  infinite $M$ data. (b) The density-density correlation $|D(r)|$ and single particle correlation $|G(r)|$, which are also fit by the power-law relations. (c) The ratio between the   magnitudes of SC pairing correlations for different bonds. The dash line indicates the spatial average over distances.  (d) The exponential fitting of the spin correlation $|S(r)|$ and $|G(r)|$.}
\label{FigS_correlations_J2_02_Jx_02_6_36}
\end{figure}

\subsection{Results of correlation functions  for additional points in SC2 phase}

We provide more evidence of the quasi-long-range SC order on $N=36\times 6$ system at $\delta = 1/12$. Figure~\ref{FigS_correlations_J2_005_Jx_005_6_36} shows various correlation functions at $J_{2}=J_{\chi }=0.05$ in the SC2 phase for an  isotropic gapped topological $d+id$-wave SC state  on $L_{y}=6$. Similar to the results at $J_{2}=0.1, J_{\chi }=0.05$, the SC pairing correlation has  power-law decaying behavior  in the infinite $M$ limit, with the Luttinger exponent  $K_{SC}\approx 1.04$. 
The $|D(r)|$ could be approximately fit by power-law decay with $K_{CDW}\approx 1.77 >K_{SC}$ and $|D(r)|$ decays faster than power-law  at long distances, 
confirming that the SC pairing correlation is the dominant correlation for this gapped isotropic topological  SC state.
The single particle correlation $|G(r)|$ can be fit only by  an exponential decay with a  short  correlation length $\xi_G\approx 2.41$.
Similarly, the spin correlation  $|S(r)|$ has an exponential decay with a short correlation length $\xi_S\approx 2.06$.

As a comparison, we also show  an anisotropic topological SC state at $J_{2}=J_{\chi }=0.1$ in the  SC2 phase in Fig.~\ref{FigS_correlations_J2_01_Jx_01_6_36}. Similar to the results at $J_{2}=0.1, J_{\chi }=0.05$, the SC pairing correlation has  power-law decaying behavior  in the infinite $M$ limit, with the Luttinger exponent  $K_{SC}\approx 0.79$. The $|D(r)|$ has power-law decay relation with $K_{CDW}>K_{SC}$.
The single particle correlation $|G(r)|$ can be fit either by a power-law or an exponential decay with a relatively long correlation length $\xi_G\approx 7.21$.
The spin correlation  $|S(r)|$ has an exponential decay with a short correlation length $\xi_S\approx 1.72$.

\subsection{Results of correlation functions in SC0 phase}

 The correlation functions at larger $J_{2}=0.2$ in the SC0 phase is given in Figs.~\ref{FigS_correlations_J2_02_Jx_02_6_36}(a) and (b).
 Both SC pairing correlations and density-density correlations follow the power-law behavior, with $K_{SC}\approx 1.10$ much smaller than
  $K_{CDW}\approx 2.02$. The ratio between the  magnitudes of SC pairing correlations for different bonds exhibit large spatial oscillations. As shown in Fig.~\ref{FigS_correlations_J2_02_Jx_02_6_36}(c), the maximum deviation from the average value of the ratio is around 0.52 at $J_{2}=0.2,J_{\chi }=0.2$. In addition, the spin correlation $|S(r)|$ has a relatively large correlation length of $\xi _{S}=5.90$ as shown in Fig.~\ref{FigS_correlations_J2_02_Jx_02_6_36}(d), which is slightly smaller than $L_{y}=6$. These behaviors imply the reduced robustness of the SC state at this parameter point ($J_{2}=0.2,J_{\chi }=0.2$), which may be near the phase boundary to a non-superconducting state.
 
\begin{figure}
\centering
\includegraphics[width=1\linewidth]{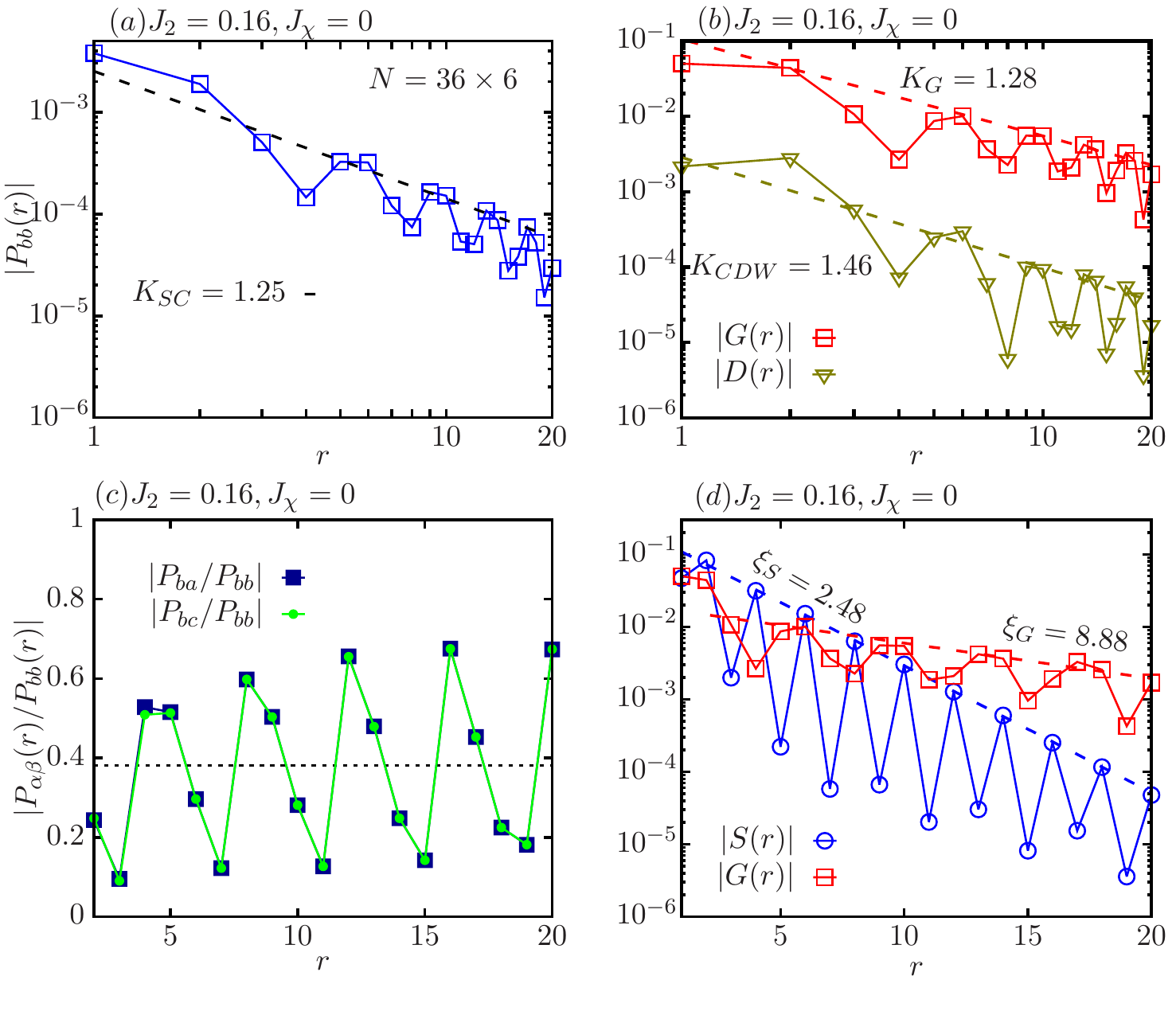}
\caption{Various correlations on $N=36\times 6$ system, obtained at $J_{2}=0.16, J_{\chi }=0$. (a) The SC pairing correlation with bond dimension $M=10000$. The Luttinger exponent is obtained by the straight-line  fitting in the log-log plot. (b) The density-density correlation $|D(r)|$ and single particle correlation $|G(r)|$, which are also fit by the power-law relations. (c) The ratio between the magnitudes of SC pairing correlations for different bonds. The dash line indicates the spatial average over distances.  (d) The exponential fitting of the spin correlation $|S(r)|$ and $|G(r)|$.}
\label{FigS_correlations_J2_016_Jx_0_6_36}
\end{figure}

\begin{figure}
\centering
\includegraphics[width=1\linewidth]{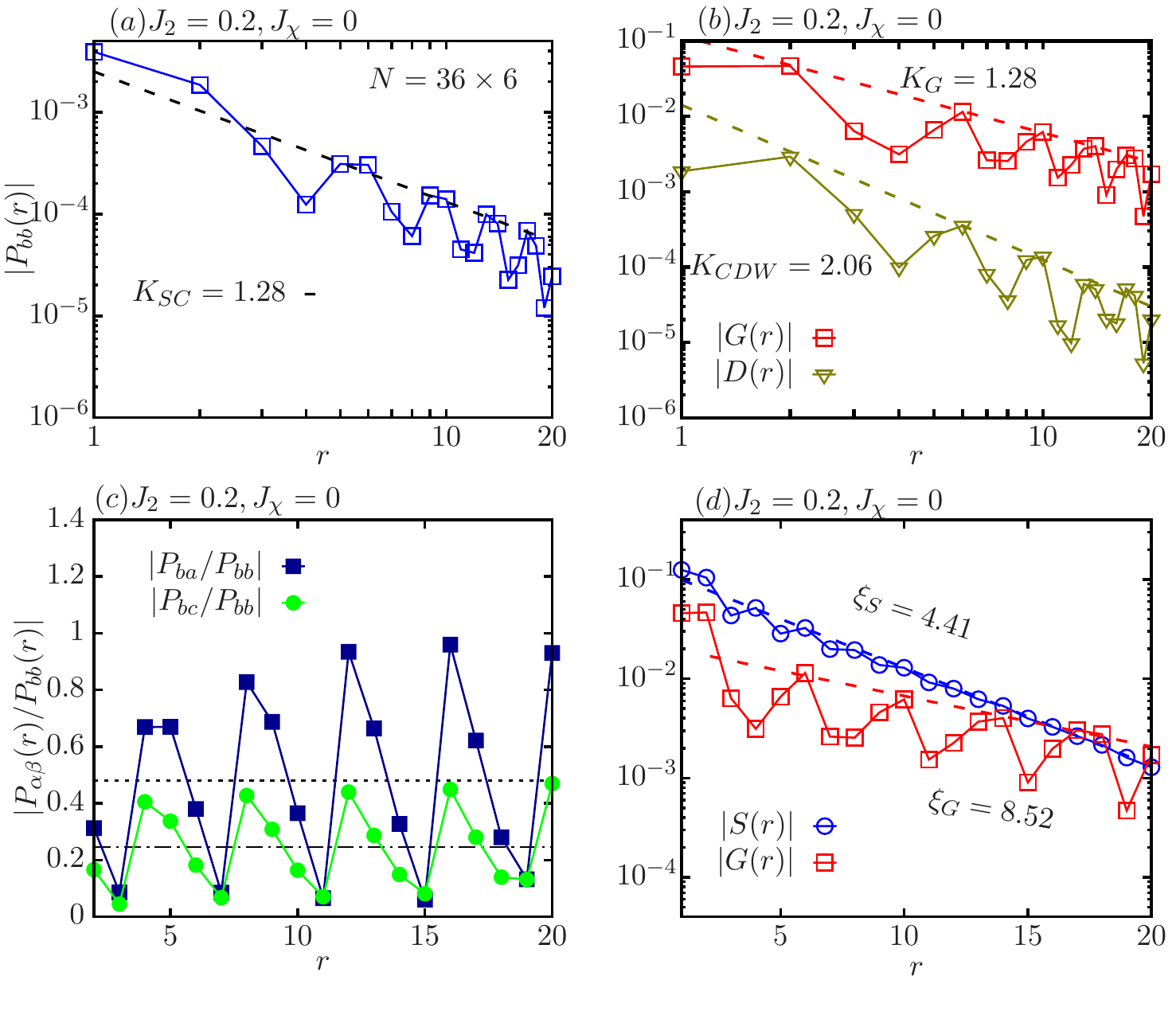}
\caption{Various correlations on $N=36\times 6$ system, obtained at $J_{2}=0.2, J_{\chi }=0$. (a) The SC pairing correlation with bond dimension $M=10000$. The Luttinger exponent is obtained by the straight-line  fitting in the log-log plot. (b) The density-density correlation $|D(r)|$ and single particle correlation $|G(r)|$, which are also fit by the power-law relations. (c) The ratio between the magnitudes of SC pairing correlations for different bonds. The dashed line and dashed dotted line indicate the spatial average over distances. (d) The exponential fitting of the spin correlation $|S(r)|$ and $|G(r)|$.}
\label{FigS_correlations_J2_02_Jx_0_6_36}
\end{figure}

The SC0 phase can be connected to the d-wave superconductivity by doping the $J_{1}$-$J_{2}$ model at $J_{\chi}=0$ as studied in previous work~\cite{jiang2021superconductivity}, which are also shown in Figs.~\ref{FigS_correlations_J2_016_Jx_0_6_36} and~\ref{FigS_correlations_J2_02_Jx_0_6_36} for $J_2=0.16$ and $0.20$, respectively. 
 When $J_{2}$ reaches 0.2, there is a visible change in the ratio of pairing orders on different bonds,
 which signals some changes in the  pairing symmetry with increasing $J_2$.
 
\subsection{Results of correlation functions in SC1 phase}

\begin{figure}
\centering
\includegraphics[width=1\linewidth]{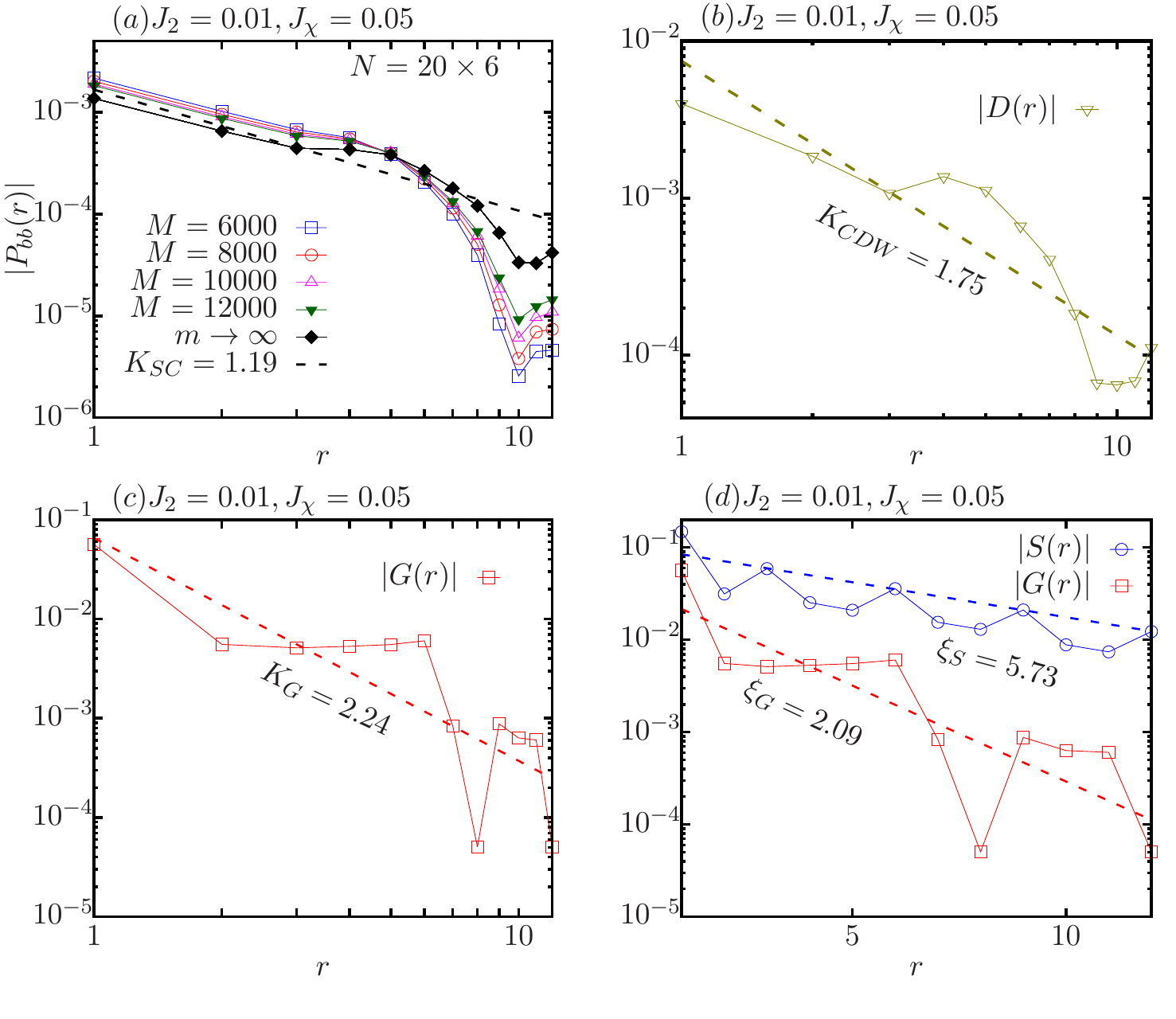}
\caption{Various correlations on $N=20\times 6$ system, obtained at $J_{2}=0.01, J_{\chi }=0.05$. (a) The SC pairing correlation with various bond dimensions. The Luttinger exponent is obtained by the straight-line fitting in the infinite bond dimension limit. (b) The density-density correlation $|D(r)|$ which is fit by the power-law relation. (c) The single particle correlation $|G(r)|$ which is also fit by the power-law relation.  (d) The exponential fitting of the spin correlation $|S(r)|$ and $|G(r)|$. The reference point is chosen at $r_{0}=(3,1)$, and the fitting is for data up to the distance $L_{x}/2$.}
\label{FigS_correlations_J2_001_Jx_005_6_20}
\end{figure}

Various correlation functions at $J_{2}=0.01, J_{\chi }=0.05$ for a system with $N=20\times 6$ are shown  in Fig.~\ref{FigS_correlations_J2_001_Jx_005_6_20}. The SC pairing correlations  $\left | P_{bb}(r) \right |$ are strongly enhanced with the increase of the bond dimension as shown in Fig.~\ref{FigS_correlations_J2_001_Jx_005_6_20}(a), where the extrapolated data for infinite bond dimension shows a possible power-law behavior with the Luttinger exponent $K_{SC}\approx 1.19$. The density-density correlations $|D(r)|$ and single particle correlation $|G(r)|$ can be  fit by the power-law relations with larger exponents $K_{CDW}\approx 1.75$ and $K_{G}\approx 2.24$, respectively, indicating dominant SC pairing correlations. The spin correlations are enhanced with a longer correlation length $\xi_s\sim L_y$. Noticing that this result is obtained on a system of a relatively short length with good convergence of the results. The full nature of the SC1 phase requires studying longer systems with larger bond dimensions due to the difficulty of the convergence for this possible critical phase.

\subsection{Comparison of spin correlations for all four phases}

\begin{figure}
\centering
\includegraphics[width=1\linewidth]{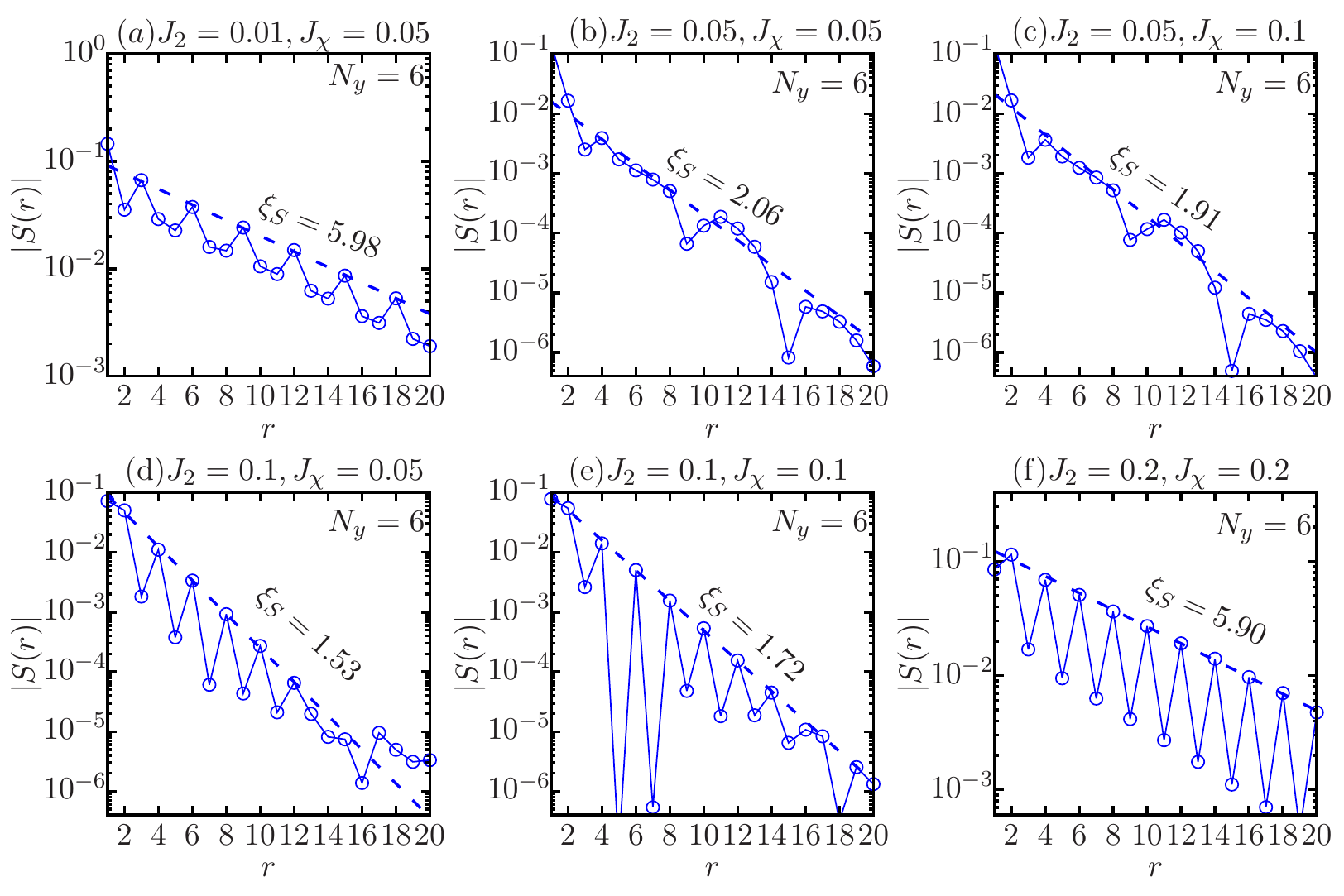}
\caption{The spin correlations for  (a) SC1 state, (b) - (c) isotropic $d+id$-wave TSC states, (d) - (e) nematic $d+id$-wave TSC states, and (f) nematic d-wave SC0 state on $N=36\times 6$ system and $\delta = 1/12$, obtained at various $J_{2}$ and $ J_{\chi }$ with infinite $M$ extrapolated data.}
\label{FigS_spin_correlation}
\end{figure}

We describe the evolution of spin correlations with different parameters of $J_{2}$ and $J_{\chi }$, as shown in Figs.~\ref{FigS_spin_correlation}(a) - (f). We find that the spin correlations always decay exponentially with a short correlation length in the SC2 regime indicating larger spin gaps crossing the whole topological superconducting phase with Chern number $C=2$. The correlation length becomes larger entering the SC1 and SC0 phases, which is consistent with the enhanced peaks in spin structure factors presented in the main text.

\subsection{Exploring possible phases in the $J_{\chi }\rightarrow 0$ limit}

\begin{figure}
\centering
\includegraphics[width=1\linewidth]{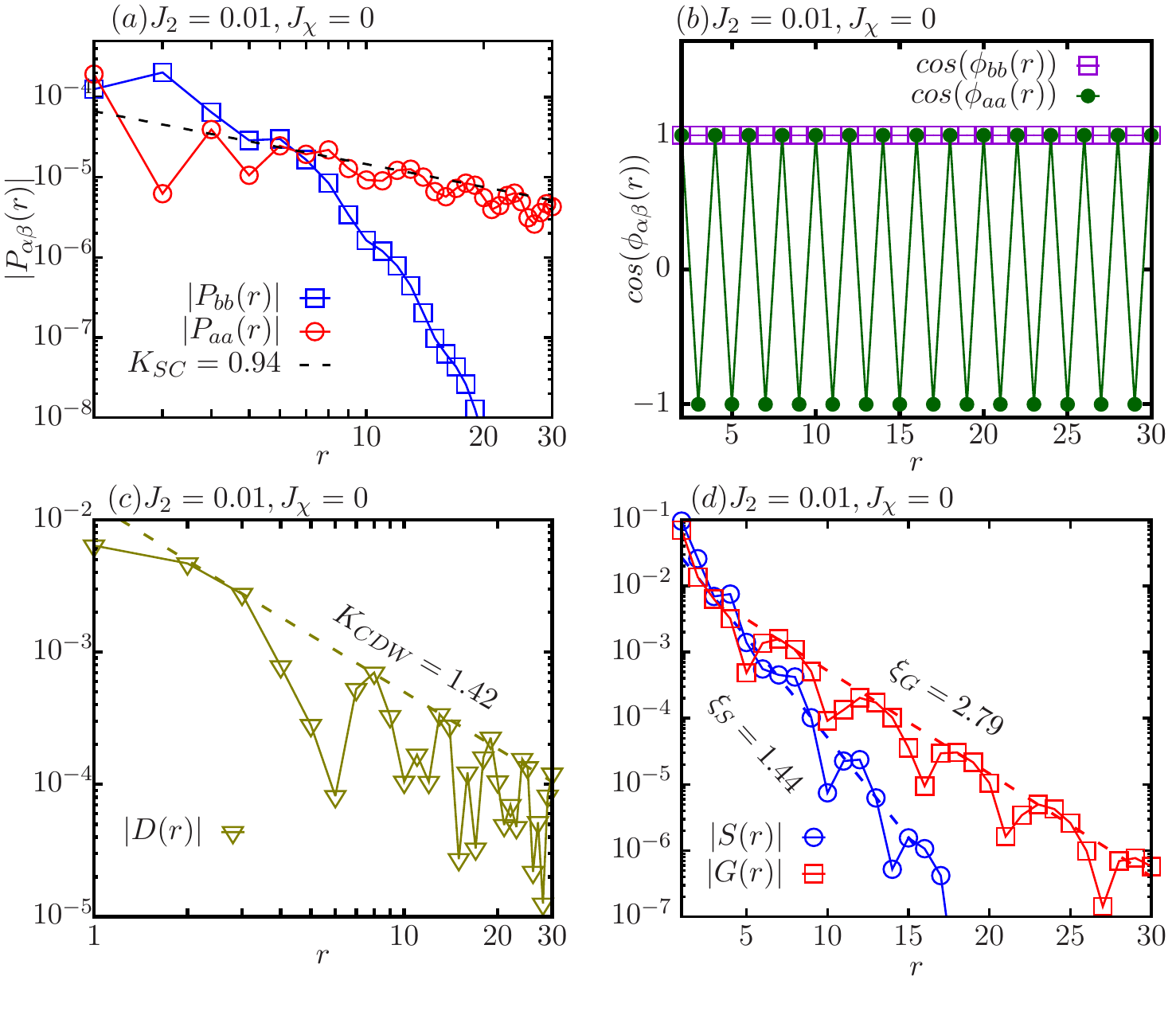}
\caption{Various correlations on $N=48\times 4$ system, obtained at $J_{2}=0.01, J_{\chi }=0$. (a) The SC pairing correlation with bond dimension $M=10000$. The Luttinger exponent is obtained by the straight-line fitting of $|P_{aa}(r)|$ in the log-log plot. (b) The phases for SC pairing correlations $\phi _{\alpha \beta}(r)$. (c) The density-density correlation $|D(r)|$, which is also fit by the power-law relation. (d) The exponential fitting of the spin correlation $|S(r)|$ and single particle correlation $|G(r)|$.}
\label{FigS_correlations_J2_001_Jx_0_4_48}
\end{figure}

There are different quantum phases  emerging in the limit of $J_{\chi }\rightarrow 0$. With a small but finite $J_{2}=0.01$ a possible pair density wave can be seen from the phases of SC pairing correlations. As shown in Fig.~\ref{FigS_correlations_J2_001_Jx_0_4_48}(a), the SC pairing correlation has a power-law decaying behavior for the a-a bond with the Luttinger exponent  $K_{SC}\approx 0.94$. The $|D(r)|$ could also be fit by power-law decay with $K_{CDW}\approx 1.42 >K_{SC}$ (Fig.~\ref{FigS_correlations_J2_001_Jx_0_4_48}(c)), while $|G(r)|$ and $|S(r)|$ has an exponential decay with short correlation lengths (Fig.~\ref{FigS_correlations_J2_001_Jx_0_4_48}(d)). This shows the dominant anisotropic SC pairing correlations. The SC pairing correlation for the b-b bond has much faster decay, and the true nature of the SC phase may require further studies on wider cylinders.

\begin{figure}
\centering
\includegraphics[width=0.83\linewidth]{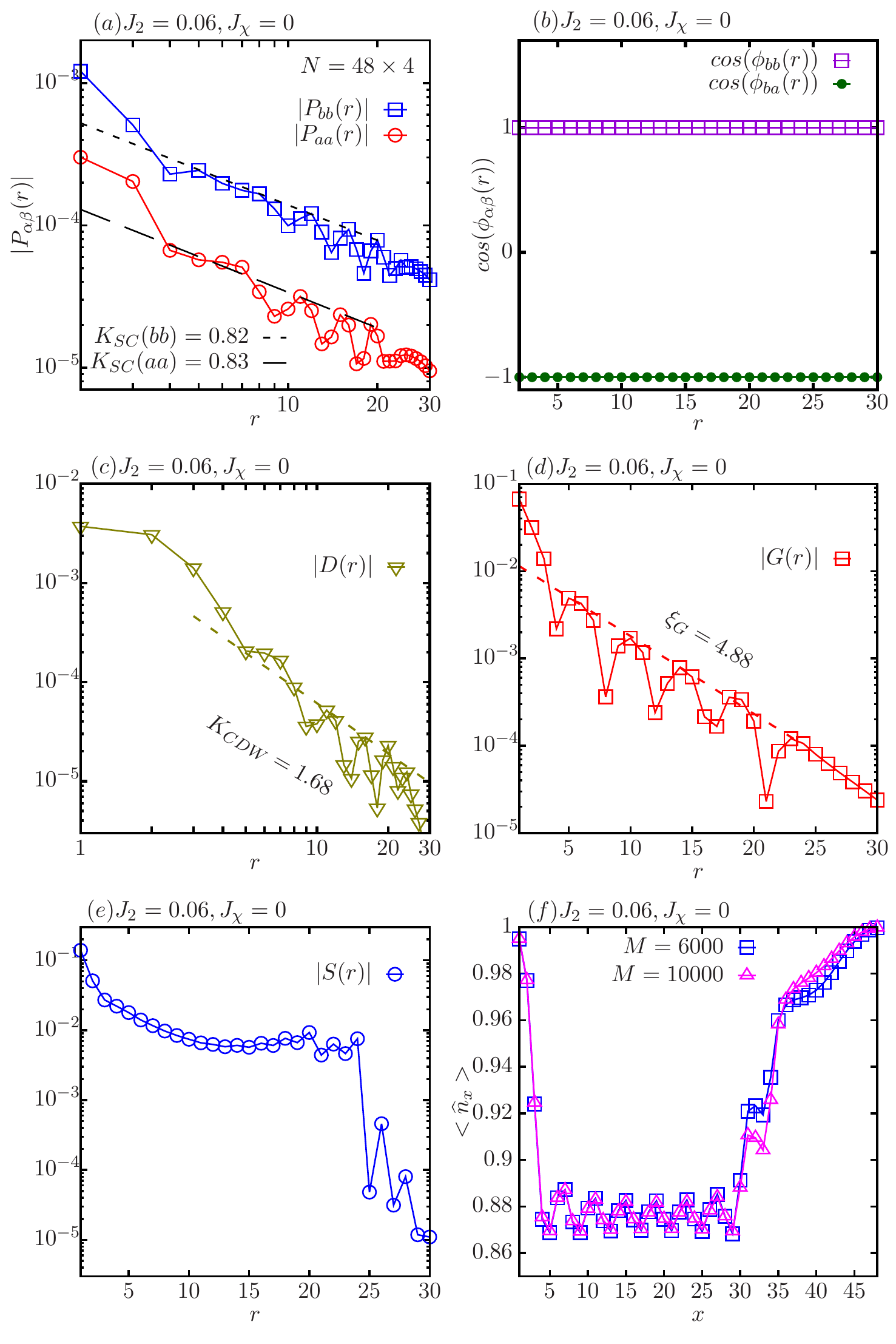}
\caption{Various correlations on $N=48\times 4$ system, obtained at $J_{2}=0.06, J_{\chi }=0$. (a) The SC pairing correlation with bond dimension $M=10000$. The Luttinger exponents are obtained by the straight-line fitting of $|P_{aa}(r)|$ and $|P_{bb}(r)|$ in the log-log plot. (b) The phases for SC pairing correlations $\phi _{\alpha \beta}(r)$. (c) The density-density correlation $|D(r)|$, which is also fit by the power-law relation. (d) The exponential fitting of single particle correlation $|G(r)|$. (e) The spin correlation $|S(r)|$. (f) The electron density obtained with different $M$. All reference points for the correlations are chosen at $x_{0}=10$}
\label{FigS_correlations_J2_006_Jx_0_4_48}
\end{figure}

As shown in Fig.~\ref{FigS_correlations_J2_001_Jx_0_4_48}(b), the phase of the SC pairing correlation for the a-a bond has an oscillation with a period of 2, indicating the pair density wave. 
When $J_{2}$ increases to 0.06, the pairing symmetry becomes consistent with $d$-wave pairing (Fig.~\ref{FigS_correlations_J2_006_Jx_0_4_48}(b)). The SC pairing correlation remains dominant comparing to $|D(r)|$ and $|G(r)|$, as shown in Figs.~\ref{FigS_correlations_J2_006_Jx_0_4_48}(c) and (d), respectively. Additionally, the SC pairing correlations for both the a-a  and b-b bonds  have power-law decaying behaviors with almost the same Luttinger exponents, as shown in Fig.~\ref{FigS_correlations_J2_006_Jx_0_4_48}(a).

However, a qualitatively different behavior is identified in the long distance for all correlation functions as shown in Figs.~\ref{FigS_correlations_J2_006_Jx_0_4_48}(a), (c), (d), and (e). Furthermore, a visible change of behavior in the electron density can be seen on the two sides of the system, as shown in Fig.~\ref{FigS_correlations_J2_006_Jx_0_4_48}(f), suggesting a phase separation that occurs around $x\approx 29$ separating hole rich region from the spin rich region in the system of $L_{x}=40$. Future studies are needed to identify the full nature of the phase with larger $L_{y}$ which may suppress the phase separation.

\section{Fermi surface evolution}
\label{SM_Fermi_surface}

\begin{figure}
\centering
\includegraphics[width=1\linewidth]{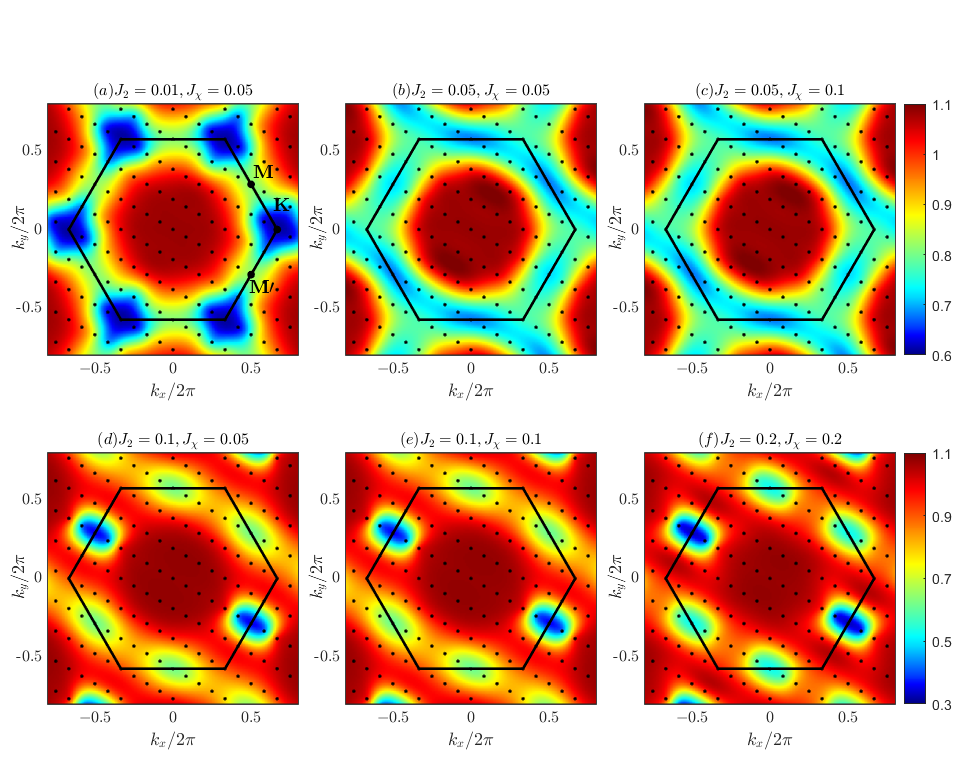}
\caption{The electron momentum distribution for (a) SC1 state, (b) - (c) isotropic $d+id$-wave TSC states, (d) - (e) nematic $d+id$-wave TSC states, and (f) nematic d-wave SC0 state  obtained at various $J_{2}$ and $ J_{\chi }$. The first Brillouin zone is indicated by the solid line with  the $\mathbf{M}$, $\mathbf{M'}$ and $\mathbf{K}$ points marked, and the black dots represent the allowed discrete momenta for the finite system with $N'=12\times6$ sites and $\delta = 1/12$ with $M=10000$.}
\label{FigS_electron_distribution}
\end{figure}

\begin{figure}
\centering
\includegraphics[width=1\linewidth]{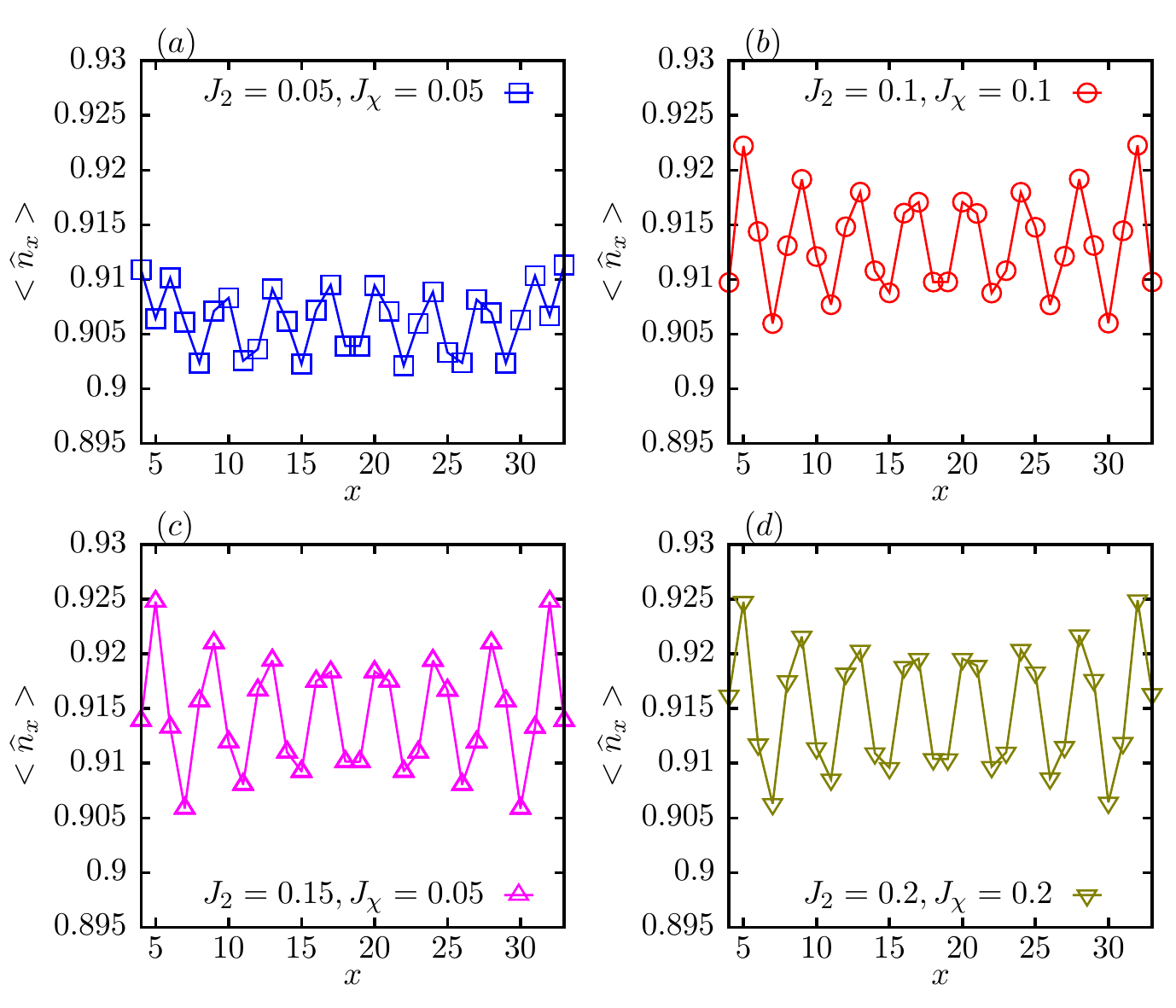}
\caption{The electron density for various $J_{2}$ and $J_{\chi}$, obtained on $N=36\times 6$ system and $\delta = 1/12$ with $M=10000$. $x$ refers to the distance from the left edge in the x-direction.}
\label{FigS_ele_density}
\end{figure}

In Figs.~\ref{FigS_electron_distribution}(a) - (f) we show the electron occupation number in the momentum space $N(\mathbf{k}) =\frac{1}{N^{'}}\sum_{i,j,\sigma}<\widehat{c}^{\dagger }_{i,\sigma }\widehat{c}_{j,\sigma }>e^{i\mathbf{k}\cdot (\mathbf{r}_{i}-\mathbf{r}_{j})}$  with summations over middle  $N'=12\times 6$ sites in a $N=36\times 6$ system for different parameters in SC1, SC2, and SC0 phases. For the SC1 phase at small $J_{2}$,  hole pockets form around the $\mathbf{K}$ points, which indicates the hole dynamics in the doped regime may be induced by  the $120^{\circ}$ AFM fluctuations. It also has a closed Fermi surface  around the $\Gamma =(0,0)$ point.  For the SC2 state  at a smaller $J_{2}=0.05$, a closed Fermi surface is formed around the $\Gamma =(0,0)$ point with near isotropic occupations in the momentum space, which is consistent with the isotropic topological $d+id$-wave SC. At larger $J_{2}=0.1$ and $0.2$, a hole pocket appears at $\mathbf{M}'$ point (which is different from the spin structure peak at $\mathbf{M}$ points), breaking the $C_3$ rotational symmetry.

\section{Electron density evolution}
\label{SM_ele_density}

As shown in Figs.~\ref{FigS_ele_density}(a) - (d), a finite electron density oscillation in real space is observed in all  states due to the tendency of charge density wave, although the density-density correlations always decay much faster than SC correlations. These results  are qualitatively  consistent with the behaviors of Luther-Emery liquid.

\section{Numerical data and program code availability}
\label{SM_data}

The main results of the study are presented within the article and its Supplementary. The digital data and the codes implementing the calculations of this study are available from the corresponding author upon reasonable request.


\end{document}